\documentclass[aps,prd,amsmath,amssymb,superscriptaddress,nofootinbib,10pt,twocolumn]{revtex4-2}

\usepackage{graphicx}
\usepackage{graphicx}
\usepackage{epstopdf}
\usepackage{hyperref}
\usepackage{color}
\usepackage{bm}
\usepackage{array}
\usepackage{booktabs}
\usepackage{float}
\usepackage{amsmath}
\usepackage{multirow}
\usepackage[utf8]{inputenc}

\allowdisplaybreaks

\definecolor{mygreen}{rgb}{0,0.5,0}  % or any shade you like

\def\apj{ApJ}
\def\apjl{ApJL}
\def\apjs{ApJS}

\def\mnras{MNRAS}
\def\physrep{PhysRep}

\def\prd{PRD}
\def\jcap{JCAP}

\begin{document}

\title{Constraining Quintessence Models with ISW\,--\,tSZ Cross-Correlations: A Comparative Analysis of Thawing, Tracker, and Scaling-Freezing Dynamics}

\author{Ayodeji Ibitoye}
\email{ayodeji.ibitoye@gtiit.edu.cn}
\affiliation{Department of Physics, Guangdong Technion - Israel Institute of Technology, Shantou, Guangdong 515063, P.R. China}
\affiliation{Centre for Space Research, North-West University, Potchefstroom 2520, South Africa}
\affiliation{Department of Physics and Electronics, Adekunle Ajasin University, P. M. B. 001, Akungba-Akoko, Ondo State, Nigeria}

\author{Shiriny Akthar}
%\email{shirinyakthar@gmail.com}
\affiliation{Department of Astronomy, Astrophysics and Space Engineering, Indian Institute of Technology Indore, Indore 453552, India}

\author{Md. Wali Hossain}
%\email{mhossain@jmi.ac.in}
\affiliation{Department of Physics, Jamia Millia Islamia, New Delhi, 110025, India}

\author{Amare Abebe}
\affiliation{Centre for Space Research, North-West University, Potchefstroom 2520, South Africa}
\affiliation{National Institute for Theoretical and Computational Sciences (NITheCS), South Africa}

\author{Prabhakar Tiwari}
\email{prabhakar.tiwari@gtiit.edu.cn}
\affiliation{Department of Physics, Guangdong Technion - Israel Institute of Technology, Shantou, Guangdong 515063, P.R. China}

\author{Xuelei Chen}
\affiliation{National Astronomical Observatories, Chinese Academy of Sciences, 20A Datun Road, Chaoyang District, Beijing 100101, P. R. China}

\author{Jackson Said}
\affiliation{Institute of Space Sciences and Astronomy, University of Malta, Malta, MSD 2080}
\affiliation{Department of Physics, University of Malta, Msida}

\author{Jacob Oloketuyi}
\affiliation{School of Physical Science and Technology, Southwest Jiaotong University, 999 Xi’an Road, Pidu District, Chengdu, 611756, Sichuan, China}
\affiliation{Department of Physics, Bamidele Olumilua University of Education, Science and Technology, Ikere-Ekiti City, P.O. Box 250, Ekiti State, Nigeria}

\begin{abstract}
We present constraints on quintessence dark energy models using the observational detection of the Integrated Sachs-Wolfe (ISW)--thermal Sunyaev-Zeldovich (tSZ) cross-correlation dataset. Our analysis compares three classes of quintessence dynamics: thawing, tracker, and scaling-freezing with the standard $\Lambda$CDM cosmology. Through a comprehensive likelihood analysis, we derive best-fit values and 68\% confidence intervals for key cosmological parameters, finding $\Omega_{\rm m} = 0.322^{+0.027}_{-0.030}$ and $\sigma_8 = 0.735^{+0.045}_{-0.035}$ for $\Lambda$CDM, with deviations in alternative models consistent within $1\sigma$. For the thawing model, we consider an exponential potential with slope $\lambda = 0.736^{+0.270}_{-0.227}$, while for the tracker and scaling-freezing models, we use inverse axion-like and double exponential potentials, respectively. Observationally, the tracker model yields $n = 5.651^{+1.625}_{-1.604}$ and $f = 0.258^{+0.149}_{-0.096}$, and the scaling-freezing model gives $\lambda_1 = 0.405^{+0.293}_{-0.322}$ and $\lambda_2 = 23.226^{+7.975}_{-7.258}$. The dimensionless tSZ amplitude ($\widetilde{W}^{\rm SZ}$) and cosmic infrared background (CIB) parameters are tightly constrained across all models, providing additional insights into astrophysical foregrounds.
Our results demonstrate the effectiveness of ISW--tSZ cross-correlations as a probe of dark energy dynamics, {with the Thawing quintessence model yielding the lowest $\chi^2_{\rm min}$ among the tested scenarios,} and highlight the need for future high-precision measurements to distinguish between quintessence models and $\Lambda$CDM.
\end{abstract}

\maketitle

\section{Introduction} 
\label{sec:introduction} 

The nature of dark energy, which constitutes approximately 70\% of the energy density of the universe, remains one of the most profound mysteries in cosmology. Since its discovery through observations of distant Type Ia supernovae \cite{Riess1998, Perlmutter1999}, dark energy has been primarily characterized by its equation of state parameter \( w \), which relates the pressure \( p \) to the energy density \( \rho \) via \( w = p/\rho \). The simplest model, the \( \Lambda \)CDM model, corresponds to \( w = -1 \), where the cosmological constant (CC), $\Lambda$, is the dark energy component. However, observations allow for the possibility of a dynamical dark energy (DDE) component, where \( w \) evolves with time \cite{Caldwell1998, Linder2003,Copeland:2006wr,Calderon_2024,lodha2024desi2024constraintsphysicsfocused,DESI:2025fii,DESI:2025zgx}.

Among the various proposals for dynamical dark energy (DDE) \cite{Copeland:2006wr,Bahamonde:2017ize}, \textit{quintessence} \cite{Ratra:1987rm,Wetterich:1987fk,wetterich1988cosmology}, a minimally coupled canonical scalar field rolling slowly at late times, has emerged as a compelling alternative to the CC. {We restrict our analysis to the non-phantom regime $w \geq -1$, a fundamental constraint imposed by the positive kinetic energy of canonical scalar fields \cite{Hossain25, Roy_Choudhury25}. This choice is further motivated by DESI Year 1 results, which reports phantom-crossing behavior at high redshifts ($z > 0.55$), but also find $w(z) \geq -1$ at low redshifts ($z < 0.55$) \cite{DESI2024} precisely the regime where the ISW--tSZ cross-correlation has peak sensitivity ($z \approx 0.45$--$0.6$) \cite{Ibitoye24}.} The evolution of quintessence models is broadly classified into three categories \cite{akthar2025generalparametrizationenergydensity}. In {\it tracker} models, the scalar field energy density decays more slowly than the background energy density, eventually dominating and driving late-time acceleration \cite{Zlatev:1998tr,Steinhardt:1999nw,Hossain25}. {\it Scaling-freezing} models {exhibit} an initial scaling {behavior}, where the scalar field energy density scales with the background \cite{Copeland:1997et}, followed by a transition to dominance due to a shallow potential or nonminimal coupling \cite{Barreiro:1999zs,Hossain:2014xha,Geng:2015fla}. In contrast, {\it thawing} models involve a field that remains frozen in the early universe, mimicking a CC, and begins evolving only recently \cite{Scherrer:2007pu}. Tracker and scaling-freezing models exhibit attractor {behavior} that helps to mitigate the cosmic coincidence problem \cite{Zlatev:1998tr,Steinhardt:1999nw}, whereas thawing dynamics is more sensitive to initial conditions.

In this work, we focus on constraining the quintessence dark energy models using the cross-correlation of large-scale structure observables, specifically the Integrated Sachs-Wolfe (ISW) effect and the thermal Sunyaev-Zel'dovich (tSZ) effect. The ISW effect \cite{Sachs1967} provides a direct signature of dark energy at low redshifts by capturing the change in the energy of cosmic microwave background (CMB) photons as they traverse time-evolving gravitational potentials \cite{Creque-Sarbinowski:2016wue,Schaefer:2008qs,Dent:2008ek,Kable:2021yws,Yengejeh:2022tpa,PhysRevD.81.083514,Vagnozzi:2021gjh}. In \cite{Ibitoye24}, the underlying ISW effect was successfully measured through its cross-correlation with tSZ data, yielding a $3.6\sigma$ detection. The thermal Sunyaev-Zel'dovich (tSZ) effect, which arises from the inverse Compton scattering of CMB photons off hot electrons in galaxy clusters, serves as a complementary probe of large-scale structure \cite{Sunyaev1972}. By combining the ISW and tSZ maps from the \textit{Planck} satellite, we can extract valuable information on the growth of cosmic structures and the evolution of dark energy \cite{Planck2015-parameter, Planck2015-XXIV}. Modeling the cross-correlation dataset with a quintessence model provides an innovative method to estimate the baryon-to-matter density ratio, i.e. \(\Omega_b/\Omega_m\), where \(\Omega_b\) denotes the normalized baryonic energy density and \(\Omega_m\) the normalized total matter density. Quintessence fields, characterized by a dynamical equation of state, affect the growth of cosmic structures and the relative distribution of matter components \cite{Zlatev1999, Wang2000}.

This model's flexibility in describing the evolution of dark energy allows for more accurate predictions of the matter power spectrum and its components, including the baryon density fraction. By leveraging the cross-correlation of ISW and tSZ effects, we can derive constraints on \(\Omega_b/\Omega_m\), offering insights into the baryonic matter's role in cosmic evolution and structure formation. This approach provides a novel and independent method to constrain \(\Omega_b\), complementing existing measurements from the CMB and Baryon Acoustic Oscillations (BAO) \cite{Planck2018TT, Alam2017}. By providing a robust test of quintessence models, this work contributes to the broader effort of refining cosmological parameters and advancing our knowledge of the universe's fundamental components.

In this study, we {utilize} our robust detection of the ISW\,--\,tSZ cross-correlation from \cite{Ibitoye24} as the primary dataset to test the quintessence model and derive constraints on the baryon density fraction alongside other cosmological and model parameters. The paper is {organized} as follows: Section~\ref{sec:dynamics} describes the models used. Section~\ref{sec:PS} discusses the matter power spectrum. Section~\ref{sec:ISWtSZ} presents the ISW--tSZ power spectrum. Section~\ref{sec:results} details the likelihood method and results. Section~\ref{sec:conclusions} summarizes our conclusions. Throughout, we assume a spatially flat FLRW cosmology, fixing $n_{\rm s}=0.965$, $\tau=0.0540$, and $\ln(10^{10}A_{\rm s})=3.043$ \cite{Planck2018TT}.

\begin{figure*}[ht]
\centering
\includegraphics[scale=0.43]{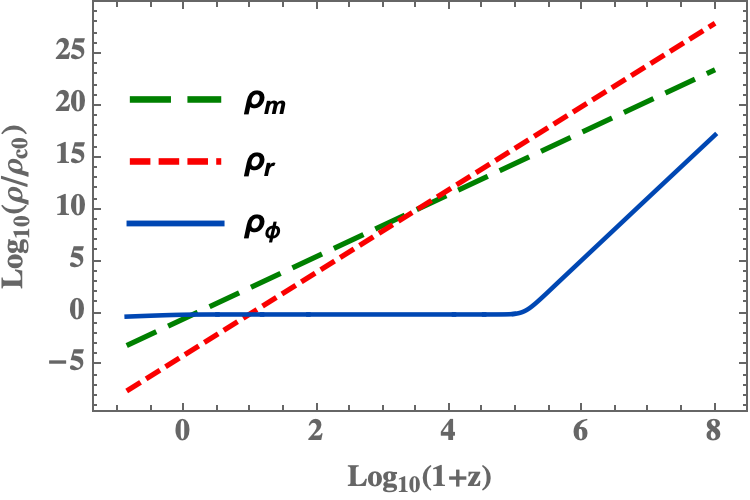} 
\includegraphics[scale=0.43]{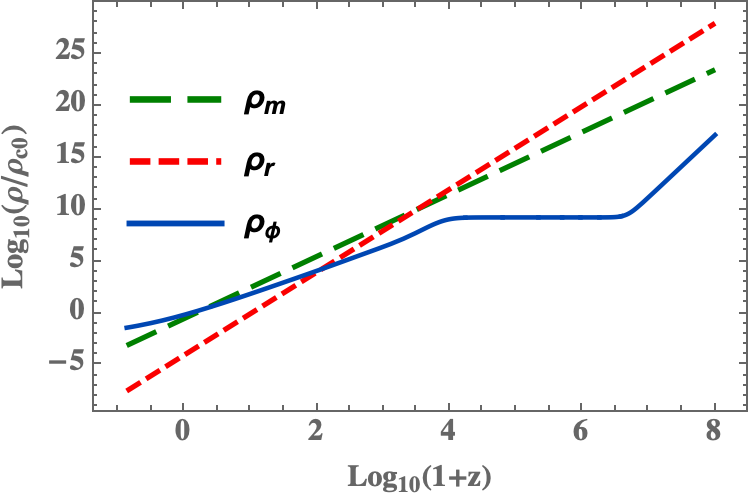}
\includegraphics[scale=0.43]{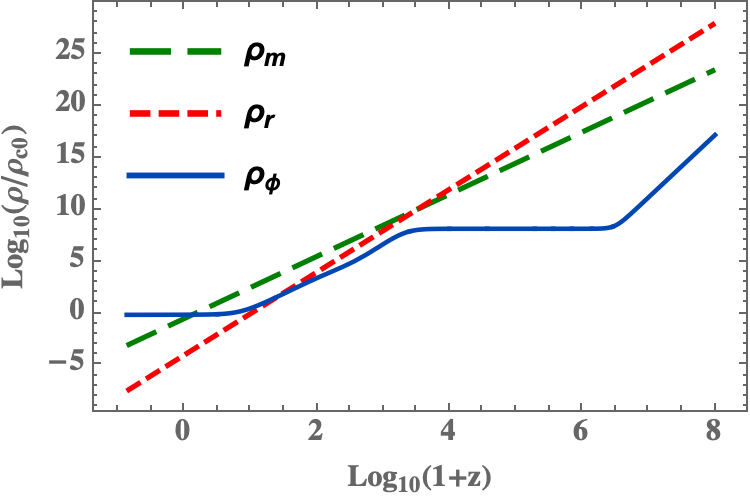}
\caption{Different scalar field dynamics has been shown. Left figure shows the thawing dynamics while middle and the right figures show the tracker and scaling-freezing dynamics, respectively.}
\label{fig:dynamics}
\end{figure*}

%%%%%%%%%%%%%%%%%%%%%%%%%%%%%%%%%%%%%%

\section{The Models}
\label{sec:dynamics}

We consider the minimally coupled canonical scalar field{ }($\phi$) governed by the following action:
\begin{equation}
    \mathcal{S} = \int d^4x \, \sqrt{-g} \left[ \frac{M_{\text{Pl}}^2}{2} R - \frac{1}{2} \partial^\mu \phi \partial_\mu \phi - V(\phi) \right] + \mathcal{S}_m + \mathcal{S}_r
    \label{action}
\end{equation}
Here, \( M_{\text{Pl}} = \frac{1}{\sqrt{8\pi G}} \) is the reduced Planck mass and {$V(\phi)$} represents the scalar field potential. $\mathcal{S}_m$ and $\mathcal{S}_r$ {represent} the action for matter and radiation, respectively. Varying the action (\ref{action}) with respect to the metric \( g_{\mu\nu} \) gives Einstein’s field equation:  
\begin{equation}
    M_{\text{Pl}}^2 G_{\mu\nu} = T^{(m)}_{\mu\nu} + T^{(r)}_{\mu\nu} + T^{(\phi)}_{\mu\nu}, 
    \label{eq:EFE}
\end{equation}
where  
\begin{equation}
    T^{(\phi)}_{\mu\nu} = \phi_{;\mu} \phi_{;\nu} - \frac{1}{2} g_{\mu\nu} (\nabla \phi)^2 - g_{\mu\nu} V(\phi).
    \label{eq:tensor}
\end{equation}
The equation of motion of the scalar field can be calculated by varying the action (\ref{action}) with respect to the scalar field \( \phi \), and it is given by  
\begin{equation}
    \Box \phi - V_{\phi}(\phi) = 0, 
    \label{eq:EOM_scalar}
\end{equation}
where the subscript \( \phi \) denotes the derivative with respect to \( \phi \).

In the flat FLRW metric, the line element is given by  
\begin{equation}
    ds^2 = -dt^2 + a^2(t) \, \delta_{ij} \, dx^i dx^j.
    \label{eq:FLRW_metric}
\end{equation}
where \( a(t) \) is the scale factor. The Friedmann equations are given by  
\begin{equation}
    3H^2 M_{\text{Pl}}^2 = \rho_m + \rho_r + \frac{1}{2} \dot{\phi}^2 + V(\phi),
    \label{eq:Friedmann1}
\end{equation}
\begin{equation}
    \left( 2\dot{H} + 3H^2 \right) M_{\text{Pl}}^2 = -\frac{1}{3} \rho_r - \frac{1}{2} \dot{\phi}^2 + V(\phi).
    \label{eq:Friedmann2}
\end{equation}
The equation of motion of the scalar field is given by  
\begin{equation}
    \ddot{\phi} + 3H \dot{\phi} + \frac{dV}{d\phi} = 0.
    \label{eq:scalar_EOM}
\end{equation}
The effective equation of state (EoS) is given by  
\begin{equation}
    w_{\text{eff}} = - \left( 1 + \frac{2}{3} \frac{\dot{H}}{H^2} \right).
    \label{eq:EoS_eff}
\end{equation}
The scalar field energy density $\rho_{\phi}$ and pressure $p_{\phi}$ {are} given by:
\begin{eqnarray}
 &&\rho_\phi=\frac{1}{2}\dot{\phi}^2+V(\phi)  \;,\\  
  &&  p_\phi=\frac{1}{2}\dot{\phi}^2-V(\phi)\;.
\end{eqnarray}
The EoS for the scalar field is therefore:
\begin{equation}
    w_{\phi}=\frac{\frac{1}{2}\dot{\phi}^2-V(\phi)}{\frac{1}{2}\dot{\phi}^2+V(\phi)}\;.
\end{equation}

For a quintessence field, the EoS ($w_\phi$) can vary from $1$ to $-1$ depending upon the relative values of the kinetic and potential terms. Hence, the scalar field energy density varies as $\rho_\phi\sim a^{-n}$, where $0\leq n \leq 6$. When $n=0$, the scalar field behaves like a CC {\it i.e.}, the potential {completely dominates the kinetic term}; and for $n=6$, the kinetic term completely dominates over the potential term. This wide variation in the evolution of $\rho_\phi$ can be classified into three classes: Thawing, Tracker, and Scaling-Freezing. Fig.~\ref{fig:dynamics} shows the different dynamics of a scalar field in a flat FLRW background. The left figure {depicts thawing dynamics}, where the scalar field remains frozen for most of the universe’s history and starts evolving in the recent past, giving rise to dynamics distinct from a CC. The middle figure represents tracker dynamics, in which the scalar field has a frozen period in the past and, once it exits this frozen period, almost—but not completely follows the background. In fact, $\rho_\phi$ decays slightly slower than the background, which eventually makes it the dominant component. The right figure shows scaling-freezing dynamics.
{For an exponential (EXP) potential with a steep slope, $\rho_\phi$ completely scales with} the background after the past frozen period. This is known as scaling {behavior} \cite{Copeland:1997et}. This is an attractor solution that makes the scalar field follow the background indefinitely. This dynamics cannot give rise to late-time acceleration. However, if we adopt a potential {\it e.g.}, a double exponential (DEXP) potential, where the potential behaves like a steep exponential in the past and becomes shallow in the recent epoch, this change in slope can give rise to dynamics that feature scaling {behavior} in the past followed by slow-roll evolution in recent times. This is known as scaling-freezing dynamics.

Here, we {consider} three distinct potentials representing the three different classes of dynamics mentioned above: an EXP potential for thawing dynamics \cite{Copeland:1997et,Scherrer:2007pu,akthar2025generalparametrizationenergydensity}, an inverse axion-like (IAX) potential \cite{Hossain_2024} for tracker dynamics, and a DEXP potential \cite{barreiro2000quintessence,J_rv_2004} for scaling-freezing dynamics:
\setlength{\abovedisplayskip}{5pt} % Adjust space before equations
\setlength{\belowdisplayskip}{5pt} % Adjust space after equations
\begin{equation}
V(\phi) = 
\left\{
\begin{array}{ll}
V_0 e^{-\lambda \phi} & \text{(EXP)} \\[6pt]
V_0 \left(1 - \cos\left(\dfrac{\phi}{f}\right)\right)^{-n} & \text{(IAX)} \\[6pt]
V_1 e^{-\lambda_1 \phi} + V_2 e^{-\lambda_2 \phi} & \text{(DEXP)}
\end{array}
\right.
\end{equation}

In all cases, $V_0, V_1, V_2$ {set} the energy scales of the potentials. In the EXP potential, $\lambda$ determines the slope of the potential and controls how quickly the field thaws at late times. For tracker dynamics, $f$ and $n$ control the tracker behavior and are treated as free parameters. For the scaling-freezing potential, $\lambda_1$ and $\lambda_2$ represent the slopes of the two exponential terms, with the first term dominating during the scaling regime and the second term becoming important at late times when the field rolls slowly. 

%%%%%%%%%%%%%%%%%%%%%%%%%%%%%%%%%
\begin{figure*}[ht!]
\centerline{\includegraphics[width=17cm]{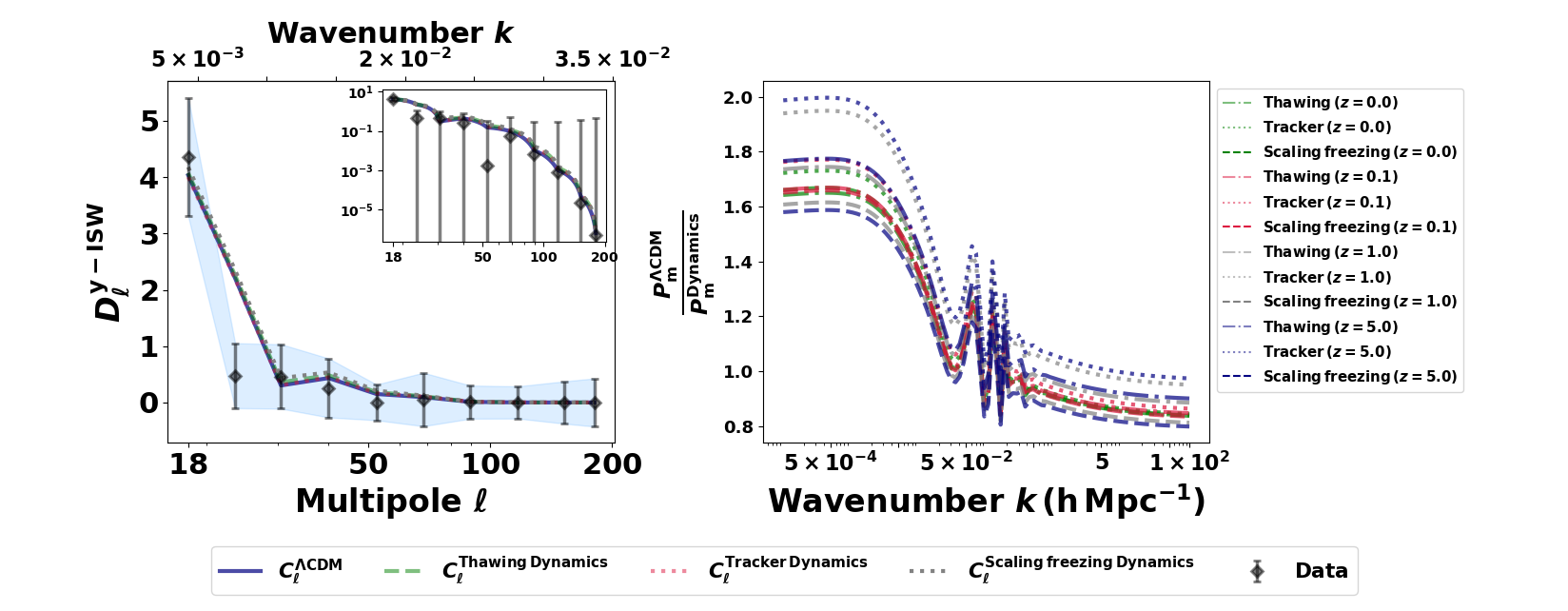}}
\caption{
{Left Panel:} Angular power spectrum of the Integrated Sachs-Wolfe (ISW)–thermal Sunyaev-Zeldovich (tSZ) effect, with uncertainties shown as an error band (dodgerblue) and error bars (grey). The data is presented in Table 1 of \cite{Ibitoye24}. The theoretical angular power spectra for various cosmological models are plotted alongside the observed data. An inset log-log plot (top-right) highlights the detailed fit of each model to the data, demonstrating that the $\Lambda$CDM model provides a better statistical fit compared to the thawing, tracker, and scaling-freezing dynamics models. $D^{\rm y-ISW}_{\ell}=10^{14} \ell(\ell+1)C^{\rm y-ISW}_{\ell}/2\pi$.
{Right Panel:} Ratios of the $\Lambda$CDM matter power spectrum ($P_{\rm m}^{\Lambda \rm CDM}$) to the matter power spectra computed for quintessence dynamics at redshifts $z = 0.0$, $z = 0.1$, $z = 1.0$, and $z = 5.0$. Different line styles and colors distinguish the thawing, tracker, and scaling-freezing dynamics models.
}
	\label{fig:Cl and Pk}
\end{figure*}

\section{Matter Power Spectrum}
\label{sec:PS}

In the sub-horizon ($k^2 \gg a^2H^2$) and quasi-static ($|\ddot{\phi}|\lesssim H|\dot{\phi}|\ll k^2|\phi|$) approximations, the evolution of the matter density contrast $\delta$ follows a second-order differential equation. Specifically, under these assumptions, the perturbations evolve according to  

\begin{equation}
    \ddot{\delta} + 2H\dot{\delta} - 4\pi G \bar{\rho}_m \delta = 0 \,,
    \label{eq:den_con_evo}
\end{equation}  
where $\bar{\rho}_m$ is the background matter density and $G$ is Newton’s gravitational constant. The solution of Eq.~\eqref{eq:den_con_evo} can be written as the superposition of growing ($D_+$) and decaying ($D_-$) modes. 

Using the growing mode $D_+$, we define the growth factor $f$ as
\begin{equation}
    f = \frac{{\rm d} \ln D_+}{{\rm d} \ln a} \, .
    \label{eq:growth}
\end{equation}
The evolution of $\sigma_8$, the root mean square amplitude of mass fluctuations within a sphere of radius $8 h^{-1}$ Mpc, with redshift is given by
\begin{equation}
    \sigma_8(z) = \sigma_8(0) \frac{D_+(z)}{D_+(0)} \, .
\end{equation}
To set the normalization, we fix the initial value of $D_+(z)$ during the matter-dominated era, which is essentially the same as in the $\Lambda$CDM model.

The power spectrum can be written as \cite{Eisenstein:1997jh,Duniya:2015nva}
\begin{equation}
    \mathcal{P}(t,k) \propto A_{\rm H}^2 \Big| D_+(a) \Big|^2 T^2(k) \left(\frac{k}{H_0}\right)^{n_s} \,.
    \label{eq:PS}
\end{equation}
Here, $A_{\rm H}$ is a normalization factor, and $n_s$ denotes the spectral index of scalar perturbations generated during inflation. The function $T(k)$ represents the transfer function, which encodes the evolution of perturbations from the primordial epoch to the late-time matter-dominated universe \cite{Eisenstein:1997ik}. Specifically, $T(k)$ relates the primordial curvature perturbation to the comoving matter perturbation. In this work, we adopt the Eisenstein–Hu fitting formula for $T(k)$ \cite{Eisenstein:1997ik}.

%------------------------------------------------------

In Figure \ref{fig:Cl and Pk}, we present the ratios of the matter power spectra:
\begin{equation}
                    \frac{P^{\Lambda{\rm CDM}}_{\rm m}(k, z^i)}{P^{\rm Dynamics}_{\rm m}(k, z^i)},
\end{equation}
evaluated at redshifts $z^i = 0.0$, $0.1$, $1.0$, and $5.0$. These results highlight the deviations of the quintessence models from the standard $\Lambda$CDM scenario across different scales and redshifts. For $\Lambda$CDM the matter power spectrum $P(k)$  is computed using the \texttt{CAMB} code \cite{Lewis2000}.

\section{ISW$-$\lowercase{t}SZ Power Spectrum} 
\label{sec:ISWtSZ}
\subsection{Theoretical Framework}
The observed temperature anisotropies on the sky can be decomposed into spherical harmonics as follows:
\begin{equation}
\Delta T(\theta, \phi) = \sum_{\ell=0}^{\infty} \sum_{m=-\ell}^{\ell} a_{\ell m} Y_{\ell m}(\theta, \phi),
\end{equation}
where $a_{\ell m}$ are the spherical harmonic coefficients, and $Y_{\ell m}(\theta, \phi)$ are the spherical harmonic basis functions.

The angular power spectrum is then estimated from these coefficients as:
\begin{equation}
C_\ell = \frac{1}{2\ell + 1} \sum_{m=-\ell}^{\ell} |a_{\ell m}|^2.
\end{equation}

Theoretically, to compute the angular power spectrum for the ISW–tSZ cross-correlation, we employ the Limber approximation, which is well-suited for small-angle analyses. While this approximation becomes less accurate at very low multipoles ($\ell$), it remains robust for $\ell \gtrsim 10$, consistent with the multipole range probed by our cross-correlation dataset. Under the Limber approximation, the angular power spectrum for the cross-correlation between two observables $A$ and $B$ (e.g., ISW and tSZ) can be expressed as:
\begin{equation}
C_\ell^{A,B} = \frac{2}{\pi} \int k^2 \, dk \, P(k) \, W_\ell^{A}(k) \, W_\ell^{B}(k),
\end{equation}
where $P(k)$ is the matter power spectrum, and $W_\ell^{A}(k)$ and $W_\ell^{B}(k)$ are the window functions associated with observables $A$ and $B$, respectively.

For the ISW–tSZ cross-correlation, the combined window function is given by:
\begin{equation}
W_\ell^{\rm ISW-tSZ}(k) = W_\ell^{\rm ISW}(k) \, W_\ell^{\rm tSZ}(k),
\end{equation}
with $W_\ell^{\rm ISW}(k)$ and $W_\ell^{\rm tSZ}(k)$ being the individual window functions for the ISW and tSZ effects, respectively. These window functions are directly related to the kernels described in Section 4.3 of \cite{Ibitoye24}.

\subsection{Observational Data}

The observed angular power spectra are computed by decomposing the full-sky ISW and Compton $y$-parameter maps into spherical harmonics. Beam effects and masking are accounted for during this process. The Limber approximation, discussed in Section 4.1, is valid for the multipole range $\ell \gtrsim 10$, ensuring consistency between the theoretical predictions and the observed data.

The observed maps are processed using the publicly available pseudo-$C_{\ell}$ MASTER algorithm implemented in the \texttt{NaMaster} package, which efficiently handles complex masks and high-resolution pixelization while mitigating mode-coupling effects introduced by the survey mask. The observed maps are decomposed into spherical harmonics to compute the angular power spectrum $C_\ell$, with corrections applied for beam effects and masking.

The covariance matrix is computed using a combination of Gaussian and non-Gaussian contributions: the Gaussian covariance is estimated from simulations using \texttt{NaMaster}, while the non-Gaussian contribution is modeled using the \texttt{pyccl} code to account for trispectrum effects, ensuring accurate treatment of uncertainties. To validate the results, {they are} compared with outputs from the {\texttt{anafast}} subroutine in the {\texttt{HEALPix}} software package and tested against Monte Carlo (MC) simulations.

For the detection, 1000 simulated ISW sky maps are generated, with 900 used to compute the covariance matrix and 100 reserved to quantify the statistical significance. The true ISW–tSZ cross-correlation is detected at $3.6\sigma$ confidence level, rejecting the no-correlation hypothesis. This comprehensive approach ensures robust characterization of the ISW–tSZ signal and its statistical significance. The autocorrelation of the Planck tSZ data is measured at a $25\sigma$ CL, while the detection of the ISW autocorrelation is still challenging due to its dominance by cosmic variance on large scales.
%%%%%%%%%%%%%%%%%%%%%%%%%%%%%%%%

%%%%%%%%%%%%%%%%%%%%%%%%%%%%%%%%%%%%%%%%%%%%%%%%%%%%%%%
\section{Likelihood Method \& Results}
\label{sec:results}
We constrain quintessence models by comparing theoretical predictions of the ISW–tSZ cross-correlation ($\mathbf{C}_{\ell}^{\rm th}(\Theta)$) with observed angular power spectra ($\mathbf{C}_{\ell}^{\rm obs}$).
The observed angular power spectra, and their associated uncertainties, as reported in Table 2 of \cite{Ibitoye22}, are binned into 10 multipole ($\ell$) bins spanning the range $\ell^{\rm min}_{\rm eff} = 18.0$ to
$\ell^{\rm max}_{\rm eff}= 181.0$. The binning process involves averaging the power spectrum values within each bin to obtain representative measurements. To perform this analysis, we define the parameter vector $\Theta$, which encompasses various categories of parameters:
\begin{equation}
\Theta = 
\left\{
\begin{aligned}
&\text{Cosmological Parameters:} && \Omega_{\rm b}, \Omega_{\rm m}, h, \sigma_8, \\
&\text{Foreground Parameters:} && B_{\rm CIB}, A_{\rm CIB}, \\
&\text{tSZ Amplitude Parameter:} && \widetilde{W}^{\rm SZ}, \\
&\text{Model Parameters:} && 
\begin{aligned}[t]
&\lambda, V_0, n, f, \\
&\lambda_1, \lambda_2, V_1, \log V_2
\end{aligned}
\end{aligned}
\right\}.
\label{eq:param}
\end{equation}
The likelihood function is constructed using the full covariance matrix ${\rm Cov}_{\rm tot}$ described in Section 2.4 of \cite{Ibitoye24}, with the $\chi^2$ statistic defined as:
\begin{equation}
\chi^2(\Theta) = \left[\mathbf{C}_{\ell}^{\rm obs} - \mathbf{C}_{\ell}^{\rm th}(\Theta)\right] {\rm Cov}_{\rm tot}^{-1} \left[\mathbf{C}_{\ell}^{\rm obs} - \mathbf{C}_{\ell}^{\rm th}(\Theta)\right]^{\rm T}.
\end{equation}

\begin{table}[h]
    \centering
    \renewcommand{\arraystretch}{1.3} % Adjust row height for better readability
    \begin{tabular}{l l c}
        \toprule
        \hline
        \textbf{Potential Type} & \textbf{Parameter} & \textbf{Prior Range} \\
        \midrule
        \multirow{2}{*}{EXP} 
            & $\lambda$ & $\mathcal{U}[0, 1.5]$ \\
            & $V_0$    & $\mathcal{U}[0, 2]$   \\
        \addlinespace
        \hline
        \multirow{3}{*}{IAX} 
            & $n$      & $\mathcal{U}[0, 10]$     \\
            & $f$     & $\mathcal{U}[0.01, 2]$   \\
            & $V_0$    & $\mathcal{U}[0, 2]$      \\
        \addlinespace
        \hline
        \multirow{4}{*}{DEXP} 
            & $\lambda_1$ & $\mathcal{U}[-1, 1]$       \\
            & $\lambda_2$ & $\mathcal{U}[0, 50]$       \\
            & $V_1$       & $\mathcal{U}[0.1, 2]$     \\
            & $\log V_2$  & $\mathcal{U}[0, 14]$      \\
            \hline
        \bottomrule
    \end{tabular}
    \caption{Prior distributions for the parameters of three quintessence potentials: EXP, IAX, and DEXP. The priors are defined as uniform distributions ($\mathcal{U}$) over specified ranges, reflecting physically motivated constraints on the model parameters. These priors ensure broad exploration of parameter space while maintaining computational feasibility.}
    \label{tab:potential_ranges}
\end{table}

\begin{table*}[ht]
    \centering
    \caption{Best-fit values and 68\% confidence limits for cosmological, astrophysical, foreground and model parameters.}
    \renewcommand{\arraystretch}{1.3}
    \begin{tabular}{l|cccccccc}
        \hline
        Model & $\Omega_{\rm b}$ & $\Omega_{\rm m}$ & $h$ & $\sigma_8$ & $\widetilde{W}^{\rm SZ}$ & $B_{\rm CIB} \, [\times 10^{-4}]$ & $A_{\rm CIB}$ & $\lambda$ \\
        \hline
        Thawing & $0.049^{+0.003}_{-0.003}$ & $0.277^{+0.032}_{-0.027}$ & $0.694^{+0.023}_{-0.018}$ & $0.746^{+0.042}_{-0.041}$ & $3.522^{+0.336}_{-0.284}$ & $5.986^{+1.806}_{-1.658}$ & $0.863^{+0.046}_{-0.041}$ & $0.736^{+0.270}_{-0.227}$ \\
        Tracker & $0.046^{+0.003}_{-0.002}$ & $0.352^{+0.015}_{-0.015}$ & $0.671^{+0.010}_{-0.008}$ & $0.712^{+0.039}_{-0.028}$ & $3.861^{+0.216}_{-0.222}$ & $3.184^{+1.236}_{-1.065}$ & $0.815^{+0.042}_{-0.045}$ & - \\
        Scaling-Freezing & $0.048^{+0.002}_{-0.003}$ & $0.301^{+0.029}_{-0.026}$ & $0.691^{+0.021}_{-0.019}$ & $0.729^{+0.045}_{-0.037}$ & $3.713^{+0.391}_{-0.338}$ & $4.540^{+1.706}_{-1.546}$ & $0.854^{+0.044}_{-0.043}$ & - \\
        ${\Lambda{\rm CDM}}$ & $0.048^{+0.003}_{-0.003}$ & $0.322^{+0.027}_{-0.030}$ & $0.676^{+0.015}_{-0.012}$ & $0.735^{+0.045}_{-0.035}$ & $2.934^{+0.377}_{-0.370}$ & $2.815^{+1.088}_{-1.059}$ & $0.900^{+0.080}_{-0.069}$ & - \\
        \hline
        Model & $V_0$ & $\rm{n}$ & $\rm{f}$ & $\lambda_1$ & $\lambda_2$ & $V_1$ & $\log V_2$ &  \\
        \hline
        Thawing & $0.170^{+0.155}_{-0.121}$ & - & - & - & - & - & - &  \\
        Tracker & - & $5.651^{+1.625}_{-1.604}$ & $0.258^{+0.149}_{-0.096}$ & - & - & - & - &  \\
        Scaling-Freezing & - & - & - & $0.405^{+0.293}_{-0.322}$ & $23.226^{+7.975}_{-7.258}$ & $0.467^{+0.283}_{-0.203}$ & $7.746^{+2.385}_{-2.056}$ &  \\
        ${\Lambda{\rm CDM}}$ & - & - & - & - & - & - & - &  \\
        \hline
    \end{tabular}
    \label{tab:parameters}
\end{table*}

The likelihood is then expressed as:
$\mathcal{L}(\Theta) \sim e^{-\chi^2/2}$ while the posterior probabilities are obtained via Bayes’ theorem:
\begin{equation}
P(\Theta | d) \propto P(\Theta) \mathcal{L}(d | \Theta),
\end{equation}
with flat priors on all parameters within physically motivated ranges, as shown in Table \ref{tab:potential_ranges}. For the EXP potential, accelerated expansion can be obtained for $\lambda<\sqrt{3}$ and $V_0$ should be of similar order to the current dark energy density, $\rho_{\rm DE0}$. $\lambda=0$ can mimic the CC. For the IAX potential, $n=0$ and $V_0=\rho_{\rm DE0}$ gives a CC-like {behavior}. $n>0$ gives tracker {behavior}. Since $n<0$ makes the potential oscillatory, we cannot get tracker {behavior} and hence we do not consider $n<0$. $f$ should be of the order of or less than the Planck mass, $M_{\rm pl}$. Larger values of $f$ can give rise to larger scalar field EoS at $z=0$ for a fixed $n$, which can make the models less favored by the data. For the DEXP potential, the parameters $\lambda_1$ and $V_0$ play the same {role} as the parameters of the EXP potential. But the parameters $\lambda_2$ and $V_2$ determine the dynamics during the early time, more specifically the scaling {behavior} during the intermediate stages, which requires larger values of $\lambda_2$ so that the potential behaves sufficiently steep during the intermediate stages. $V_2$ {determines} the energy scale of the early frozen period of the scalar field and the duration of the scaling regime. In general, $V_2\geq \rho_{\rm DE0}$. The CC-like {behavior} for the DEXP potential requires $\lambda_1=\lambda_2=0$, $V_1\sim\rho_{\rm DE0}$ and $V_2$ can be either zero or of similar order to $V_1$. For estimating parameters, we employ nested sampling using the \texttt{UltraNest} package~\cite{ultranest,ultranest2}, which is well-suited for exploring high-dimensional parameter spaces and estimating Bayesian evidence. The algorithm initializes live points randomly within the prior ranges and iteratively replaces the point with the lowest likelihood until convergence criteria are met. Convergence is monitored by ensuring the uncertainty on the log-evidence falls below a specified threshold (typically $< 0.1$), and that the effective sample size (ESS) exceeds 1000, ensuring robust posterior sampling. The parameter space is explored with \texttt{UltraNest}'s default settings, including adaptive refinement and dynamic updates to the proposal distribution for efficient sampling. Posterior distributions are then analyzed using \texttt{GetDist}~\cite{Antony25}, with marginalized constraints summarized in Table~\ref{tab:parameters}. Best-fit model predictions are shown in Figure~\ref{fig:Cl and Pk}, demonstrating excellent agreement between the cosmological models and observations.

%%%%%%%%%%%%%%%%%%%%%%%%%%%%%%%%%%%%%%%%%%%%%%%%%%%%%%%%%%%%%%%%%
\subsection{Impact on cosmological parameters}
\begin{figure}
\centerline{\includegraphics[width=9cm]{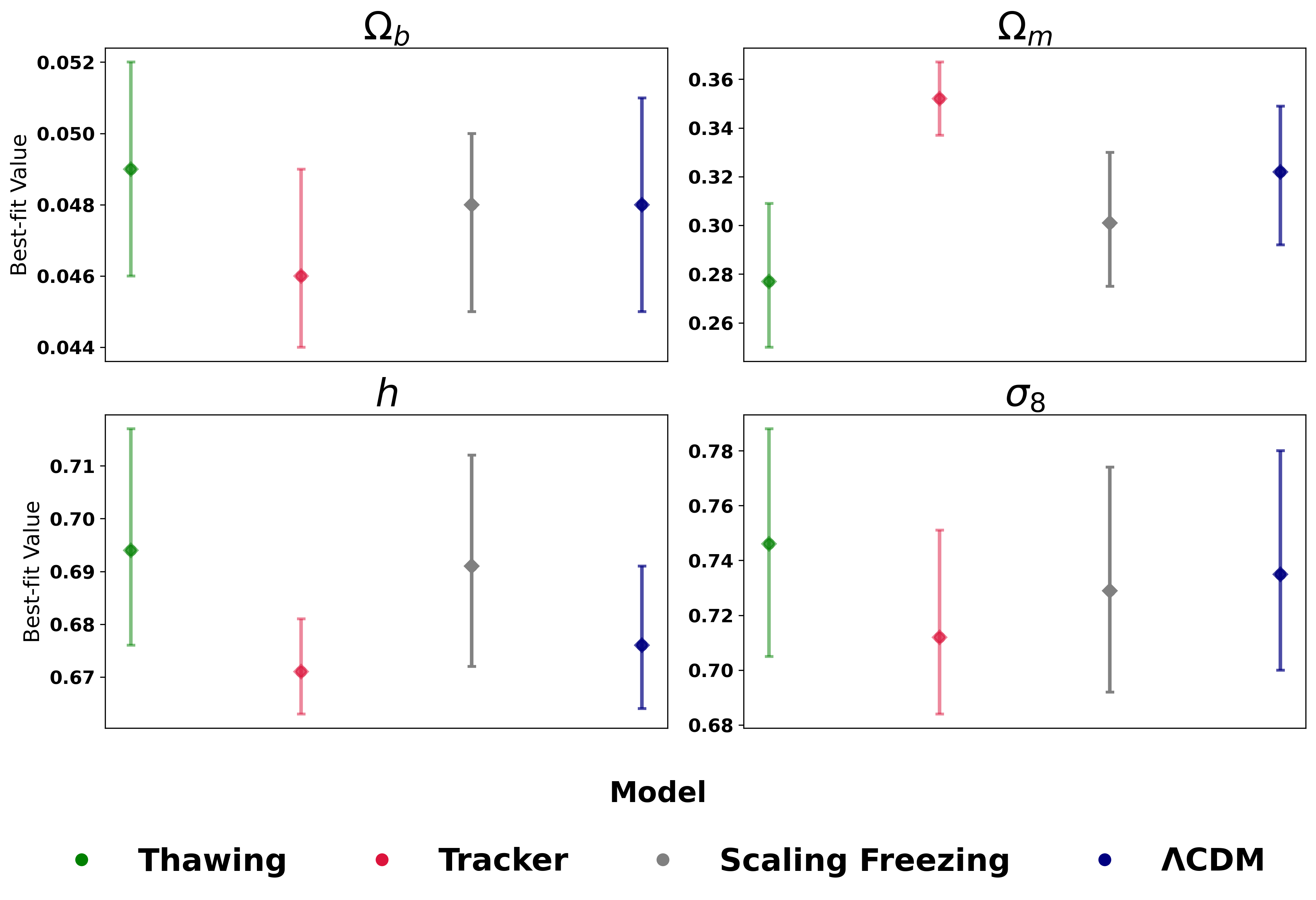}}
   \caption{Final Estimates of Model Parameters with Uncertainties. 
     Final estimates of our model parameters are computed as the marginalized 1D (M1D) values from their respective posterior distributions. 
     Each panel corresponds to an individual cosmological parameter ($\Omega_b$, $\Omega_m$, $h$, and $\sigma_8$), displaying the best-fit values grouped by the models, while error bars depict the lower and upper uncertainties associated with each estimate.}
 	\label{fig:compare_cosmological_constraints}
 \end{figure}

\begin{figure*}
\centerline{\includegraphics[width=11cm]{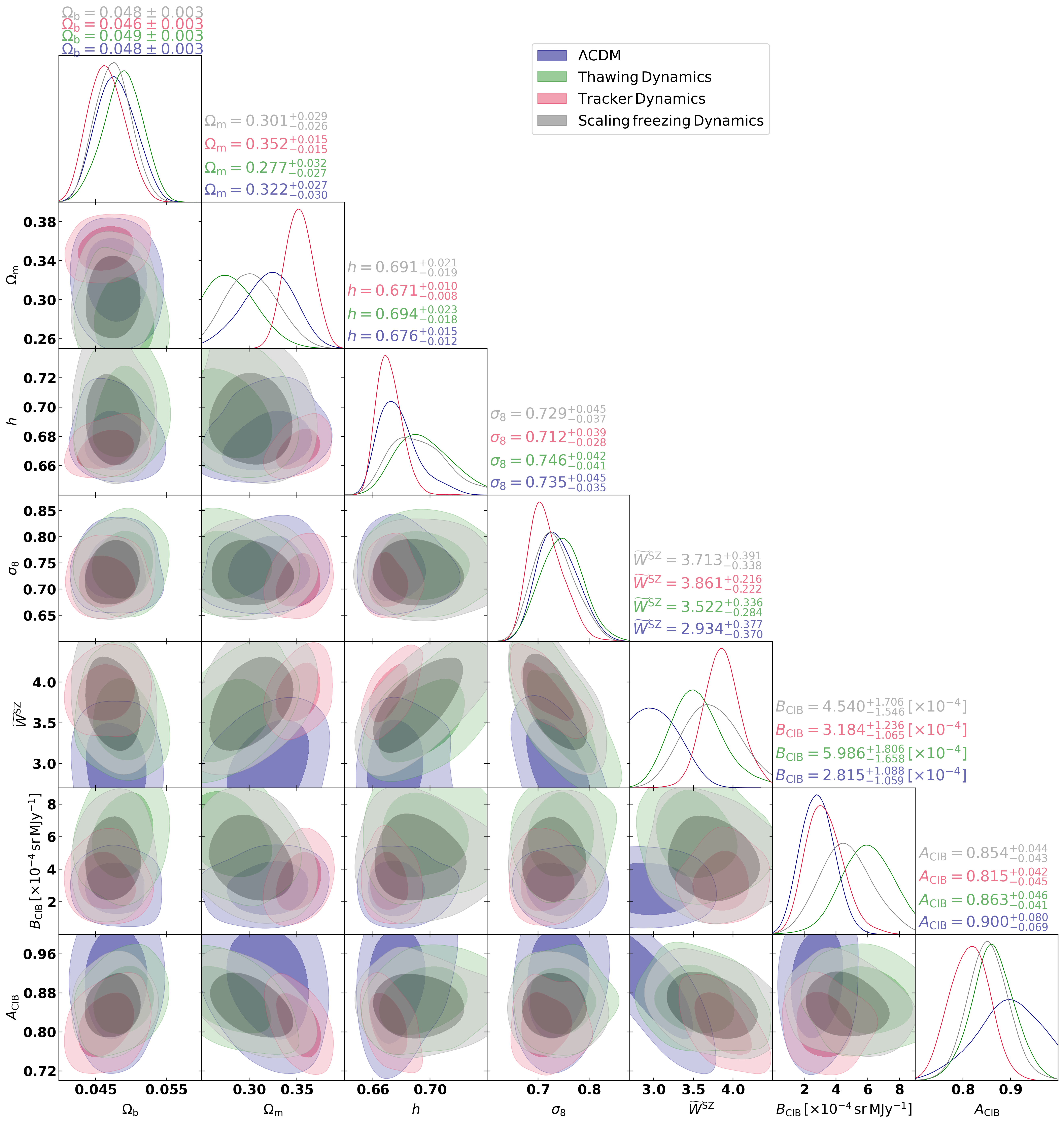}}
\caption{ Posterior distributions from MCMC with the full covariance matrix for the seven free parameters in our Quintessence model. The figures show the joint constraints
for all parameter pairs and the marginalized distributions for each parameter along the table diagonal.}
	\label{fig:compare_constraints}
\end{figure*}

We analyze how the best-fit cosmological parameters vary across different dark energy models, focusing on their consistency with the $\Lambda$CDM. In Tab.~\ref{tab:parameters}, we present the best-fit values and 68\% confidence limits on all parameters. Fig.~\ref{fig:compare_cosmological_constraints} shows the graphical representation of the best-fit value and corresponding 1$\sigma$ bound on the parameter for each scenario. Fig.~\ref{fig:compare_constraints} shows the 1$\sigma$ and 2$\sigma$ contours of the seven free parameters, cosmological, foreground and tSZ amplitude parameters listed in Eq.~\ref{eq:param}. According to Table~\ref{tab:parameters} while most parameters remain broadly consistent across models, some show mild but notable deviations, which we quantify in terms of their statistical significance ($\sigma$). For the baryon density parameter, $\Omega_{\rm b}$, the differences are relatively small: the Thawing model is consistent with $\Lambda$CDM at $0.236\sigma$, while the Tracker and Scaling-Freezing models show consistencies of $0.512\sigma$ and $0.000\sigma$, respectively. These minor variations, all well within $1\sigma$, indicate a high degree of consistency in $\Omega_{\rm b}$ across the models. This agreement suggests that $\Omega_{\rm b}$ is a robust parameter, largely insensitive to the choice of different dark energy models considered in this paper. 

In contrast, the matter density parameter, $\Omega_{\rm m}$, exhibits some variation across the models. The Thawing model deviates from $\Lambda$CDM by $1.10\sigma$, while the Tracker and Scaling-freezing models differ by $0.93\sigma$ and $0.53\sigma$, respectively. These deviations remain within the range of statistical insignificance ($<2\sigma$), indicating no strong evidence for tension between the models. However, the Tracker model predicts a slightly higher $\Omega_{\rm m}$ compared to the other models, which could reflect differences in its treatment of dark energy or the relative contributions of matter components. While these variations are not statistically significant, they highlight the sensitivity of $\Omega_{\rm m}$ to the underlying assumptions of each model and emphasize the need for further investigation into these differences with more precise data.

The baryon density fraction relative to the total matter density, $\Omega_b / \Omega_m$, provides additional insight into the distribution of baryonic and dark matter across different cosmological models. The estimated values for $\Omega_b / \Omega_m$ are as follows:

\[
\Omega_b / \Omega_m =
\left\{
\begin{aligned}
    &\text{Thawing:} && 0.177^{+0.017}_{-0.019}, \\
    &\text{Tracker:} && 0.131^{+0.006}_{-0.005}, \\
    &\text{Scaling-Freezing:} && 0.160^{+0.014}_{-0.016}, \\
    &\Lambda\text{CDM:} && 0.149^{+0.012}_{-0.011}.
\end{aligned}
\right.
\]
These results highlight variations in $\Omega_b / \Omega_m$ across the models, reflecting differences in their underlying assumptions about the interplay between baryonic and dark matter components. Notably, the Thawing model predicts a higher baryon fraction compared to $\Lambda$CDM, while the Tracker model yields the lowest value. This underscores the sensitivity of the baryon-to-matter ratio to the treatment of dark energy and matter dynamics.

\begin{figure*}[t]
\centerline{\includegraphics[width=13cm]{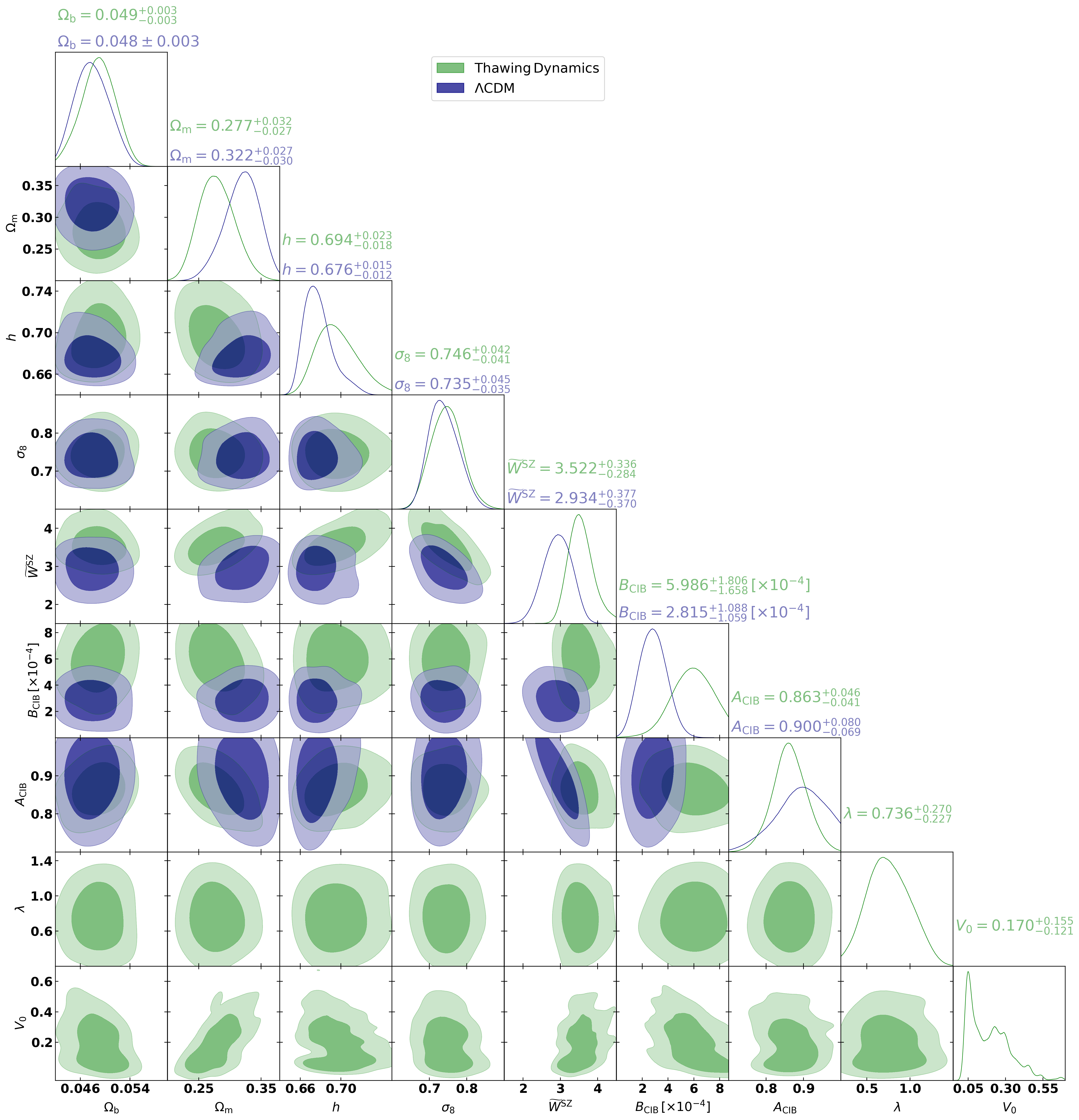}}
\caption{ The posterior distributions and correlations of key cosmological parameters from Thawing Dynamics of Quintessence Models Compared to the $\Lambda{\rm CDM}$ Framework. The constraints includes the Thawing dynamics sensitive parameters. The filled contours represent the 68\% and 95\% confidence regions for the Thawing Dynamics (green) and $\Lambda{\rm CDM}$ (navy) models, illustrating the differences in parameter estimates and their uncertainties, as well as the impact of different dynamical behaviors on cosmological constraints.}
	\label{fig:Thawing_and_LCDM_constraints}
\end{figure*}

We now turn to the constraints on the Hubble constant, $h$ across the different models. The Thawing model is consistent with $\Lambda$CDM at $0.73\sigma$, while the Tracker and Scaling-Freezing models show consistencies of $0.31\sigma$ and $0.62\sigma$, respectively. These small deviations, all within $1\sigma$, suggest that the expansion rate of the universe remains largely unaffected by the choice of the model. This agreement reinforces the robustness of $h$ as a cosmological parameter and supports its reliability in cosmological studies.

For the amplitude of matter fluctuations, $\sigma_8$, the differences between the models are small but noticeable. The Thawing model is consistent with $\Lambda$CDM at $0.20\sigma$, while the Tracker and Scaling-Freezing models show consistencies of $0.40\sigma$ and $0.11\sigma$, respectively. These deviations remain well within the range of statistical insignificance ($<2\sigma$), indicating no strong evidence for tension between the models. The slightly higher $\sigma_8$ values in the Thawing model could indicate differences in the models' predictions for the linear growth of perturbations, which may arise from variations in the effective dark energy equation of state or the interaction between dark energy and matter.

The dimensionless tSZ amplitude parameter ($\widetilde{W}^{\rm SZ}$) exhibits notable sensitivity across the different dark energy models, reflecting its dependence on the underlying cosmological dynamics. For the $\Lambda{\rm CDM}$ model, $\widetilde{W}^{\rm SZ}$ is constrained to $2.934^{+0.377}_{-0.370}$, which serves as a baseline for comparison. In contrast, the Thawing, Tracker, and Scaling-Freezing models yield higher values of $3.522^{+0.336}_{-0.284}$, $3.861^{+0.216}_{-0.222}$, and $3.713^{+0.391}_{-0.338}$, respectively. These variations are closely tied to differences in the matter density ($\Omega_{\rm m}$) and the amplitude of matter fluctuations ($\sigma_8$), as models with higher $\Omega_{\rm m}$ or $\sigma_8$ tend to produce stronger tSZ signals due to enhanced clustering of matter and hotter gas reservoirs in galaxy clusters.
%%%%%%%%%%%%%%%%%%%%%%%%%%%%%%%%

%--------------------------------------------------------
\begin{figure*}
\centerline{\includegraphics[width=13cm]{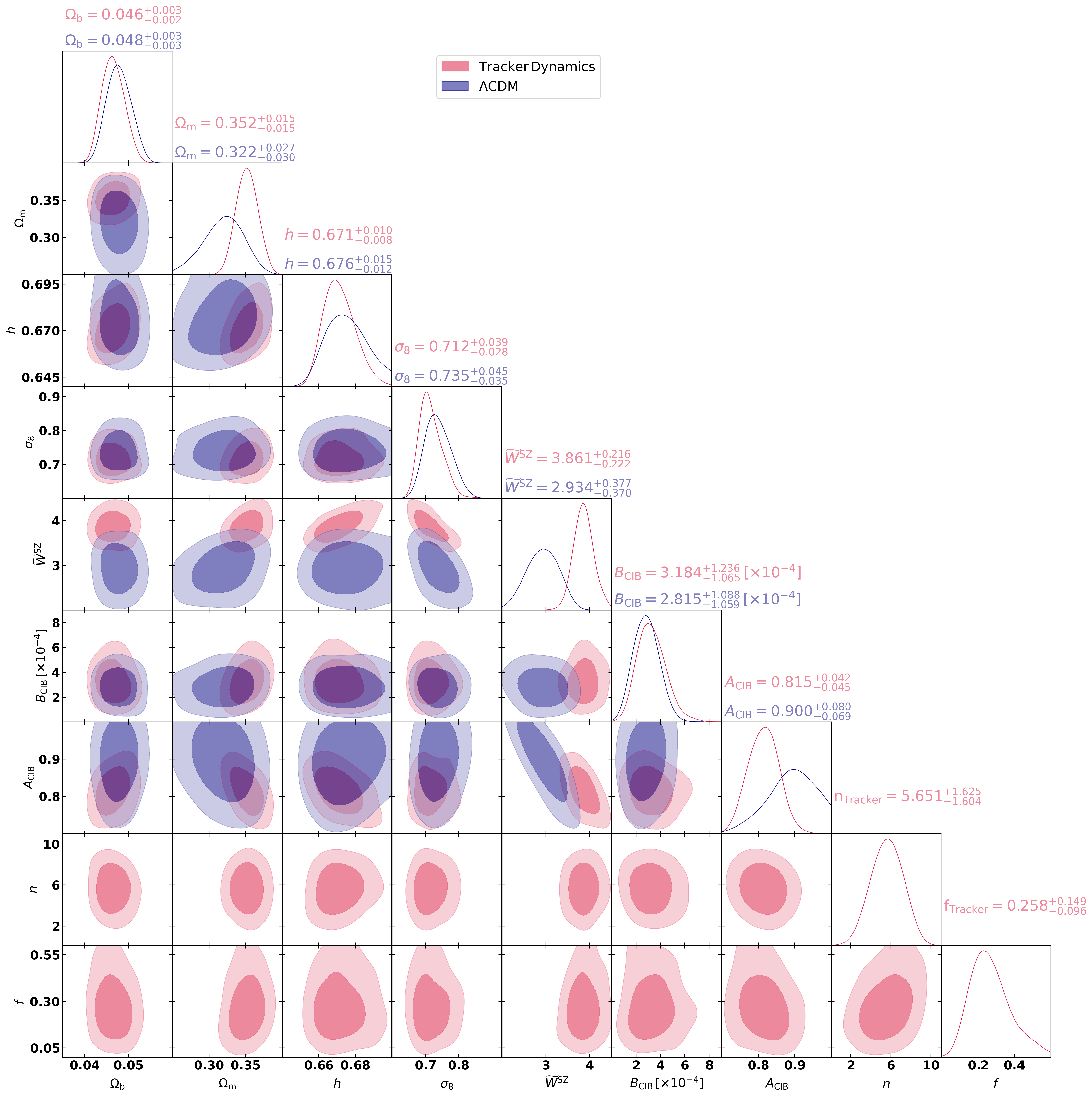}}
\caption{Exploring the Parameter Space of Tracker Dynamics in Quintessence Models Compared to the Standard Cosmological Scenario. This analysis presents the 68\% and 95\% confidence regions for Tracker Dynamics (crimson) and $\Lambda{\rm CDM}$ (navy), along with the one-dimensional distributions along the diagonal. The results reveal distinct differences in parameter estimates and uncertainties. Notably, the Tracker model indicates a higher baryon density and exhibits different behaviors in matter density and the Hubble parameter, suggesting that the dynamics of dark energy can significantly influence cosmological constraints. This study highlights the importance of considering alternative dark energy models to enhance our understanding of the universe's evolution.}
	\label{fig:Tracker_and_LCDM_constraints}
\end{figure*}
  
% - - - - - - - - - - - - - - - - - - - - - - - - - - - - 
\begin{figure*}
\centerline{\includegraphics[width=13cm]{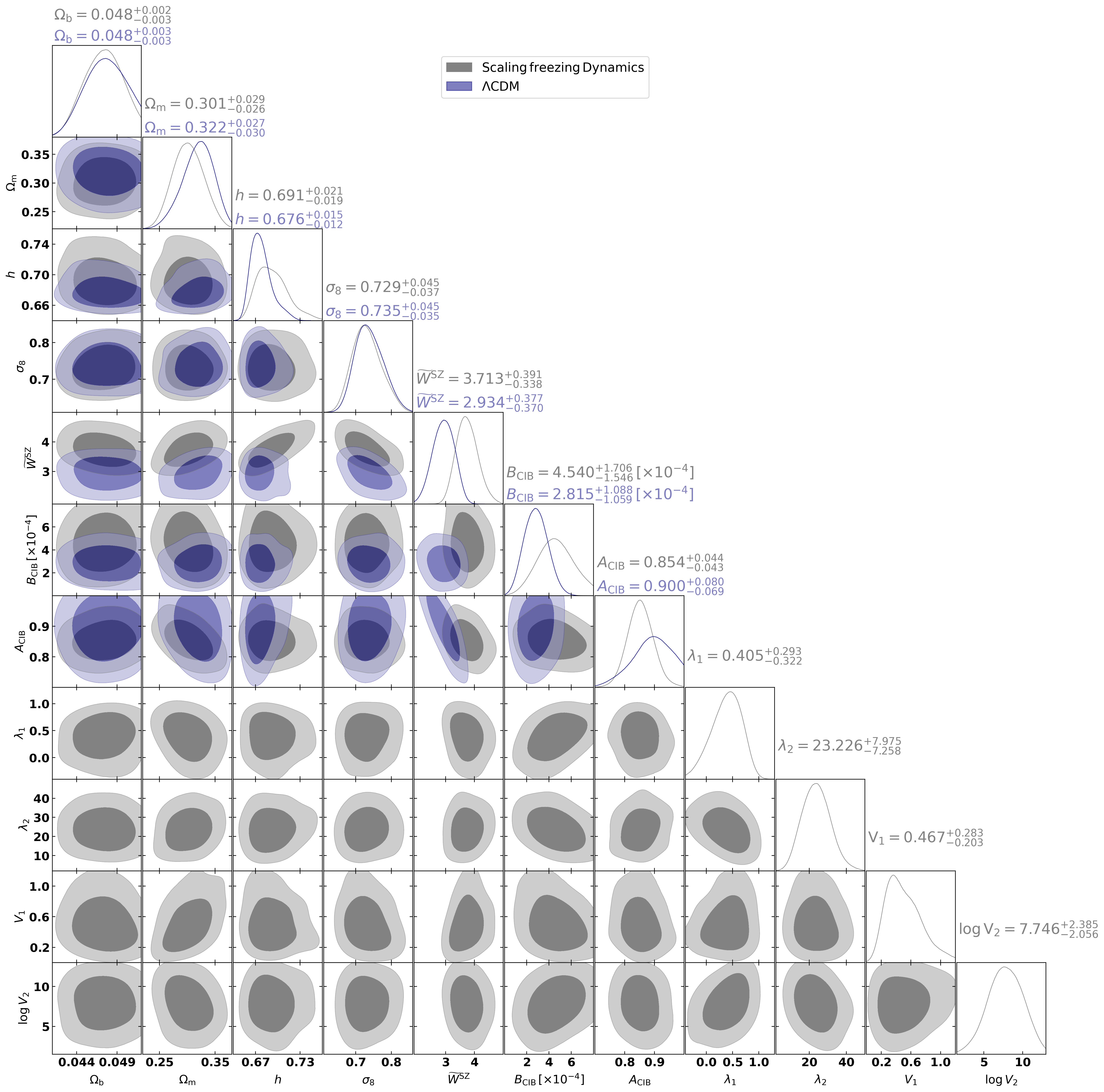}}
\caption{Comparison of Parameter Constraints from Scaling-Freezing Dynamics of Quintessence Models and the $\Lambda{\rm CDM}$ Framework. The triangle plot illustrates the posterior distributions and correlations of key cosmological parameters, including baryon density ($\Omega_{\rm b}$), matter density ($\Omega_{\rm m}$), Hubble parameter ($h$), and the amplitude of density fluctuations ($\sigma_8$). The filled contours represent the 68\% and 95\% confidence regions for the Scaling-Freezing Dynamics (grey) and $\Lambda{\rm CDM}$ (navy) models, highlighting the differences in parameter estimates and their uncertainties.}
	\label{fig:SF_and_LCDM_constraints}
\end{figure*}

\subsection{Implications for dark energy dynamics}
\label{ssec:Quintessence-Specific Parameters}

We now discuss extensively the constraints on model parameters in the Quintessence class, which reveal insights into the dynamics of dark energy. For the thawing model (Fig.~\ref{fig:Thawing_and_LCDM_constraints}), the slope of the potential $\lambda$ is constrained to $0.74^{+0.14}_{-0.16}$, indicating a dynamics explicitly different than the $\Lambda$CDM at the 1$\sigma$ level as $\lambda=0$ is outside 1$\sigma$. This result is particularly interesting as the recent BAO observation by DESI points towards a dynamical {behavior} of dark energy \cite{DESI2024,Calderon_2024,Jiang24,lodha2024desi2024constraintsphysicsfocused,DESI:2025zpo,DESI:2025zgx}. {
It is worth noting that the ISW--tSZ cross-correlation signal is dominated by low multipoles ($\ell \lesssim 50$), as shown in Figure~\ref{fig:Cl and Pk}, and is primarily sensitive to low redshifts in this regime (see bottom panel of Figure 6 in our detection analysis \cite{Ibitoye24}). For these scales, the redshift kernel peaks at $z \approx 0.45$--$0.6$, with the majority of its weight concentrated at $z < 1.0$. This range overlaps with the low-redshift regime ($z < 0.55$) where DESI Year 1 finds $w(z) \geq -1$ \cite{DESI2024}. Consequently, our preference for Thawing quintessence ($w > -1$) is consistent with DESI’s findings in this overlapping epoch. Rather than being in tension, the two probes are complementary: ISW--tSZ provides an independent evidence of non-phantom dark energy dynamics where DESI also supports $w \geq -1$.
} 
{
Canonical quintessence models are restricted to $w \geq -1$ and cannot reproduce the phantom-crossing behavior ($w < -1$) reported in some DESI analyses. Our constraints therefore apply only to dynamical dark energy within the non-phantom regime.
} 
The other parameter, $V_0$, is consistent with low-energy scales ($V_0 = 0.17 \pm 0.10$). In the tracker model (Fig.~\ref{fig:Tracker_and_LCDM_constraints}), we find $n = 5.57^{+1.37}_{-1.42}$ and $f = 0.27^{+0.15}_{-0.09}$, also suggesting a significant departure from $\Lambda{\rm CDM}$ dynamics. {
However, given the modest significance ($3.6\sigma$) of the ISW--tSZ detection, the preference for $n \approx 5.6$ in the Tracker model should be interpreted as indicative rather than conclusive, pending validation from higher-precision data.
} For the scaling-freezing model, the scaling-freezing dynamics with DEXP potential parameters ($\lambda_1, \lambda_2, V_1, \log V_2$) are tightly constraint (Fig.~\ref{fig:SF_and_LCDM_constraints}), with $\lambda_2 = 23.3^{+6.23}_{-5.61}$ and $\log V_2 = 7.93 \pm 1.80$. These values highlight the role of early-time scaling behavior in shaping the late-time evolution of the field. Having discussed the constraints on model parameters, we show in Figures~\ref{fig:Thawing_and_LCDM_constraints}, \ref{fig:Tracker_and_LCDM_constraints}, and \ref{fig:SF_and_LCDM_constraints}, the complete posterior distributions of the parameter spaces and their correlations. 

To further explore the implications of these constraints, we analyze the Integrated Sachs-Wolfe (ISW) effect. We find that the Scaling Freezing model exhibits the strongest ISW signal with an ISW strength of $0.270$, followed by the Thawing model ($0.224$), $\Lambda$CDM ($0.205$), and the Tracker model ($0.197$). The higher ISW strength in the Scaling Freezing model can be attributed to its more pronounced evolution of the gravitational potential during the matter-to-dark-energy transition, reflecting a more rapid change in the underlying dark energy dynamics compared to the other models. This result suggests that dark energy models characterized by significant deviations from $\Lambda$CDM behavior, particularly those with steep transitions in their equation of state, are likely to produce stronger ISW effects. Such findings have important implications for understanding the nature of dark energy, as they highlight the sensitivity of the ISW effect to the dynamical properties of dark energy.

{
The ISW--tSZ cross-correlation probes the large-scale, quasi-linear decay of gravitational potentials, statistically averaged over the full sky. In contrast, the so-called ISW "tension" arises from local, nonlinear measurements using stacked cosmic superstructures \cite{Granett2008, Ili2014, Cai2014}. While these approaches are complementary, they probe different aspects of structure growth and are subject to distinct systematics.} Among our models, Scaling-Freezing yields the strongest ISW signal (0.270), followed by Thawing (0.224), $\Lambda$CDM (0.205), and Tracker (0.197). The Scaling-Freezing model is statistically disfavored by our ISW--tSZ data ($\Delta{\rm BIC} = +3.36$), while the preferred Thawing model produces a modest enhancement relative to $\Lambda$CDM.

Future observations aimed at improving the precision of ISW measurements, such as deeper surveys of large-scale structure and high-resolution cosmic microwave background data, could provide critical tests for distinguishing between competing dark energy models and further constraining their parameter spaces. In addition to this, recent work by \cite{Dong25} reports the second model-dependent measurement of the gravitational potential decay rate (DR) using DESI DR9 galaxy samples. Their analysis achieves a detection significance of $3.1\sigma$ across the redshift range $0.2 \leq z < 1.4$, highlighting the growing importance of DR as a cosmological observable. Future work could incorporate DR measurements alongside ISW analyses to further refine constraints on both constant and time-varying dark energy models. These efforts would not only enhance our understanding of the late-time expansion history of the universe but also shed light on the fundamental physics governing dark energy.

\subsection{Comparison of Cosmological Models}
\begin{figure}
    \centerline{\includegraphics[width=7cm]{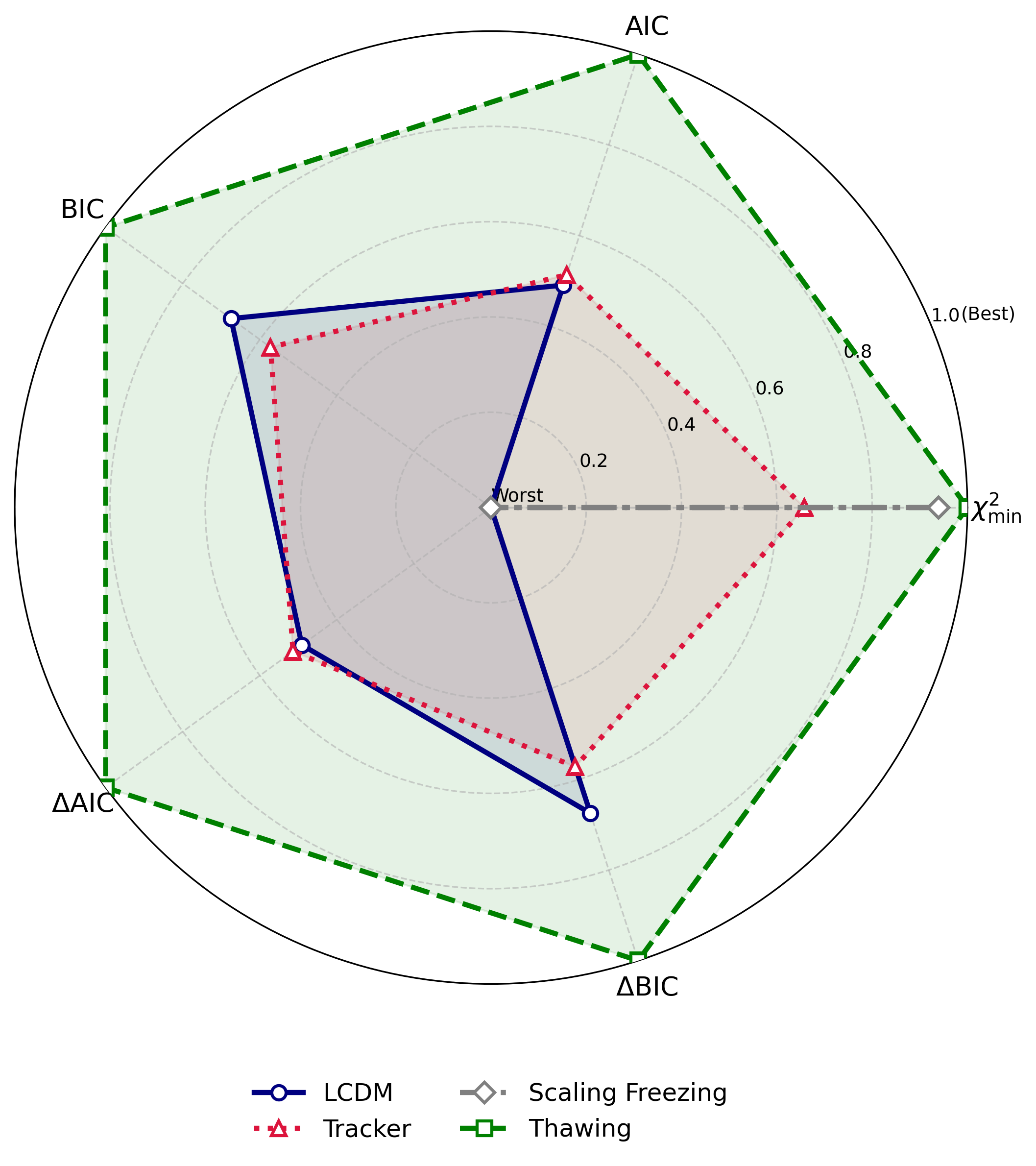}}
\caption{Radar chart comparing the statistical performance of four cosmological models: $\Lambda$CDM, Thawing, Tracker, and Scaling-Freezing. Performance is evaluated using five metrics: $\chi^2_{\rm min}$, AIC/BIC (information criteria balancing fit and parsimony), and their relative differences to $\Lambda$CDM ($\Delta$AIC, $\Delta$BIC). All metrics follow a "lower-is-better" paradigm and are independently normalized to [0,1], where 1 = best and 0 = worst model for each metric. This ensures larger polygon areas correspond to superior overall performance. The Thawing model (green) dominates with the largest polygon, reflecting an optimal balance of fit and parsimony. Scaling-Freezing (darkgrey) follows, with strong $\chi^2_{\rm min}$ performance offset by lower parsimony (weak AIC/BIC scores). Tracker (crimson) and $\Lambda$CDM (navy) have progressively smaller polygons, indicating weaker overall performance. Better-performing models are plotted on top for visibility.}

\label{fig:model_comparison}
\end{figure}

\begin{table*}
    \centering
    \caption{Goodness-of-fit and information criteria for cosmological models, evaluated at maximum likelihood (ML) parameters. All metrics satisfy nested model consistency ($\chi^2_{\rm min} \leq \chi^2_{\rm min}(\Lambda{\rm CDM})$).}
    \label{tab:corrected_model_metrics}
    \begin{tabular}{lccccc}
        \toprule
        Model               & $\chi^2_{\rm min}$ (ML) & AIC   & BIC   & $\Delta \text{AIC}$ & $\Delta \text{BIC}$ \\
        \midrule
        $\Lambda$CDM        & 16.39                   & 30.39 & 32.51 & 0.00    & 0.00     \\
        Thawing             & 10.16                   & 28.16 & 30.89 & $-2.23$             & $-1.62$             \\
        Tracker             & 12.29                   & 30.29 & 33.02 & $-0.10$             & $0.51$              \\
        Scaling-Freezing    & 10.54                   & 32.54 & 35.87 & $2.15$              & $3.36$              \\
        \bottomrule
    \end{tabular}
    \label{corrected_stats}
\end{table*}

To evaluate the viability of the cosmological models under consideration, we compute the minimum chi-square ($\chi^2_{\text{min}}$), the Akaike Information Criterion (AIC), and the Bayesian Information Criterion (BIC), along with their differences relative to the standard $\Lambda{\rm CDM}$ model. These metrics jointly assess goodness-of-fit and model complexity, with lower values indicating stronger statistical support. The AIC and BIC are defined as
\begin{eqnarray}
    \text{AIC} &=& 2N + \chi^2_{\text{min}}, \nonumber \\
    \text{BIC} &=& N \ln(K) + \chi^2_{\text{min}},
\end{eqnarray}
where $N$ is the number of free parameters, $K = 10$ is the number of effective multipole bins in the ISW--tSZ data, and $\chi^2_{\text{min}}$ quantifies the fit quality. The relative differences $\Delta \text{AIC} = \text{AIC}_{\text{model}} - \text{AIC}_{\Lambda \text{CDM}}$ and $\Delta \text{BIC} = \text{BIC}_{\text{model}} - \text{BIC}_{\Lambda \text{CDM}}$ measure the evidence for or against each model compared to $\Lambda$CDM. Following standard interpretation \cite{Burnham2004}, $|\Delta| < 2$ implies substantial support, while $|\Delta| > 10$ indicates decisive disfavor.

Critically, all $\chi^2_{\text{min}}$ values are evaluated at the \textit{maximum likelihood point} of the posterior distribution, ensuring statistically valid comparison across nested model families. As theoretically required, both the Thawing and Scaling-Freezing models extensions of $\Lambda$CDM that recover it in the limit of vanishing field dynamics yield $\chi^2_{\text{min}} \leq \chi^2_{\text{min}}(\Lambda{\rm CDM})$, showing proper sampling of the $\Lambda$CDM boundary and the validity of our inference.

{
Table~\ref{tab:corrected_model_metrics} shows that the Thawing model achieves the lowest $\chi^2_{\rm min} = 10.16$, compared to $\Lambda$CDM ($\chi^2_{\rm min} = 16.39$). It also has the lowest AIC (28.16) and BIC (30.89), giving $\Delta{\rm BIC} = -1.62$ relative to $\Lambda$CDM, which indicates substantial support under standard criteria \cite{Burnham2004}. The Scaling-Freezing model has a comparable $\chi^2_{\rm min} = 10.54$ but higher BIC (35.87) due to its larger parameter count ($N=11$), yielding $\Delta{\rm BIC} = +3.36$. The Tracker model gives $\chi^2_{\rm min} = 12.29$ and $\Delta{\rm BIC} = +0.51$. The best-fit (maximum-likelihood) Thawing slope is $\lambda = 0.974$, distinct from the $\Lambda$CDM limit ($\lambda = 0$); the Bayesian posterior median is $\lambda = 0.736^{+0.270}_{-0.227}$. All extended models satisfy $\chi^2_{\rm min} \leq \chi^2_{\rm min}(\Lambda{\rm CDM})$, as required for nested models. The relative performance of the models is visualized in Figure~\ref{fig:model_comparison}.
}

\section{Conclusion}
\label{sec:conclusions}

We have presented constraints on canonical quintessence dark energy models: Thawing, Tracker, and Scaling-Freezing using the $3.6\sigma$ ISW--tSZ cross-correlation detection from \cite{Ibitoye24}. This probe is sensitive to late-time gravitational potential decay ($z \lesssim 1$) and operates independently of early-universe physics, avoiding degeneracies present in joint CMB+BAO analyses.

{Within this dataset, the Thawing model achieves the lowest $\chi^2_{\rm min} = 10.16$ compared to $\Lambda$CDM ($\chi^2_{\rm min} = 16.39$), with $\Delta{\rm BIC} = -1.62$ and $\Delta{\rm AIC} = -2.23$, indicating substantial statistical support under standard criteria \cite{Burnham2004}. The best-fit Thawing slope is $\lambda = 0.974$, distinct from the $\Lambda$CDM limit ($\lambda = 0$). All extended models satisfy the theoretical requirement $\chi^2_{\rm min} \leq \chi^2_{\rm min}(\Lambda{\rm CDM})$ for nested scenarios.} 

{The analysis reveals a mild but persistent $\sim 1.7\sigma$ tension in $\sigma_8$, with ISW--tSZ-inferred values peaking near 0.74, compared to the \textit{Planck} 2018 value of $0.811 \pm 0.012$. This discrepancy remains after applying Gaussian Planck priors on $(\Omega_b, \Omega_m, h, \sigma_8)$, suggesting it is not an artifact of weak external constraints, but likely reflects a characteristic of the ISW–tSZ data itself.}

{Importantly, the preference for the Thawing model persists when incorporating Planck 2018 constraints on $(\Omega_b, \Omega_m, h, \sigma_8)$: the statistical support shifts from $\Delta\mathrm{BIC} = -1.62$ (without priors) to $\Delta\mathrm{BIC} = -3.83$ (with priors). Model-specific parameters show minimal shifts (Appendices~\ref{app:planck_priors_cosmo} and~\ref{app:planck_priors_dde}), indicating that the preference is not solely a consequence of weak constraints on background cosmology. However, since external priors can affect model comparison through parameter volume effects, this preference should be viewed as internal consistency within our framework, not as definitive evidence for dynamical dark energy.}

Future CMB experiments (CMB-S4, Simons Observatory)~\cite{CMB_S4, SO_25} and large-scale structure surveys (DESI, Euclid, Rubin LSST)~\cite{Euclid, Euclid-collaboration, LSST2012} will dramatically improve the precision of ISW cross-correlations. When combined with galaxy clustering, weak lensing, or supernova data, these datasets will enable robust tests of dark energy dynamics and determine whether the Thawing preference persists in a full cosmological context. {While joint analyses with BAO, SNe Ia, and CMB are essential for definitive constraints, they lie beyond the scope of this probe-focused study. We refer the reader to companion works by our co-author and collaborators: \cite{Hossain25} for tracker models, and \cite{Sohail25} for thawing and scaling-freezing dynamics, where these models are constrained using the full cosmological concordance dataset. For now, the ISW--tSZ cross-correlation serves as a complementary probe of late-time cosmic acceleration, with our analysis showing consistency with non-phantom dynamical dark energy.}

\acknowledgments

The postdoctoral funds of the Guangdong Technion-Israel Institute of Technology, China support this project. AI acknowledges the partial support of the Alliance of International Science Organizations, Grant No. ANSO-VF-2024-01. XC acknowledges support from the National Science Foundation China, Grant No. NSFC 12361141814.

\section*{Data Availability Statement}
The ISW--tSZ cross-correlation data and associated uncertainties supporting the findings of this study are published in \cite{Ibitoye24}. The underlying ISW and tSZ maps are publicly available in the \textit{Planck} Legacy Archive \cite{PlanckLegacyArchive}.

This work made use of the following software packages: {\sc HEALPix} \cite{healpix}, {\sc MASTER} \cite{Hivon02}, \texttt{NaMaster} \cite{NaMaster}, {\sc UltraNest} \cite{ultranest, ultranest2}, and {\sc GetDist} \cite{Antony25}.

\appendix

% --------------------------
% Appendix 1:Impact of Planck Gaussian Priors on Cosmological Parameter Constraints

% --------------------------
% Appendix A: Impact of Planck Gaussian Priors on Cosmological Parameter Constraints

\section{Stability of Cosmological Parameters Under External Constraints}\label{app:planck_priors_cosmo}

\begin{figure*}
    \centerline{\includegraphics[width=13cm]{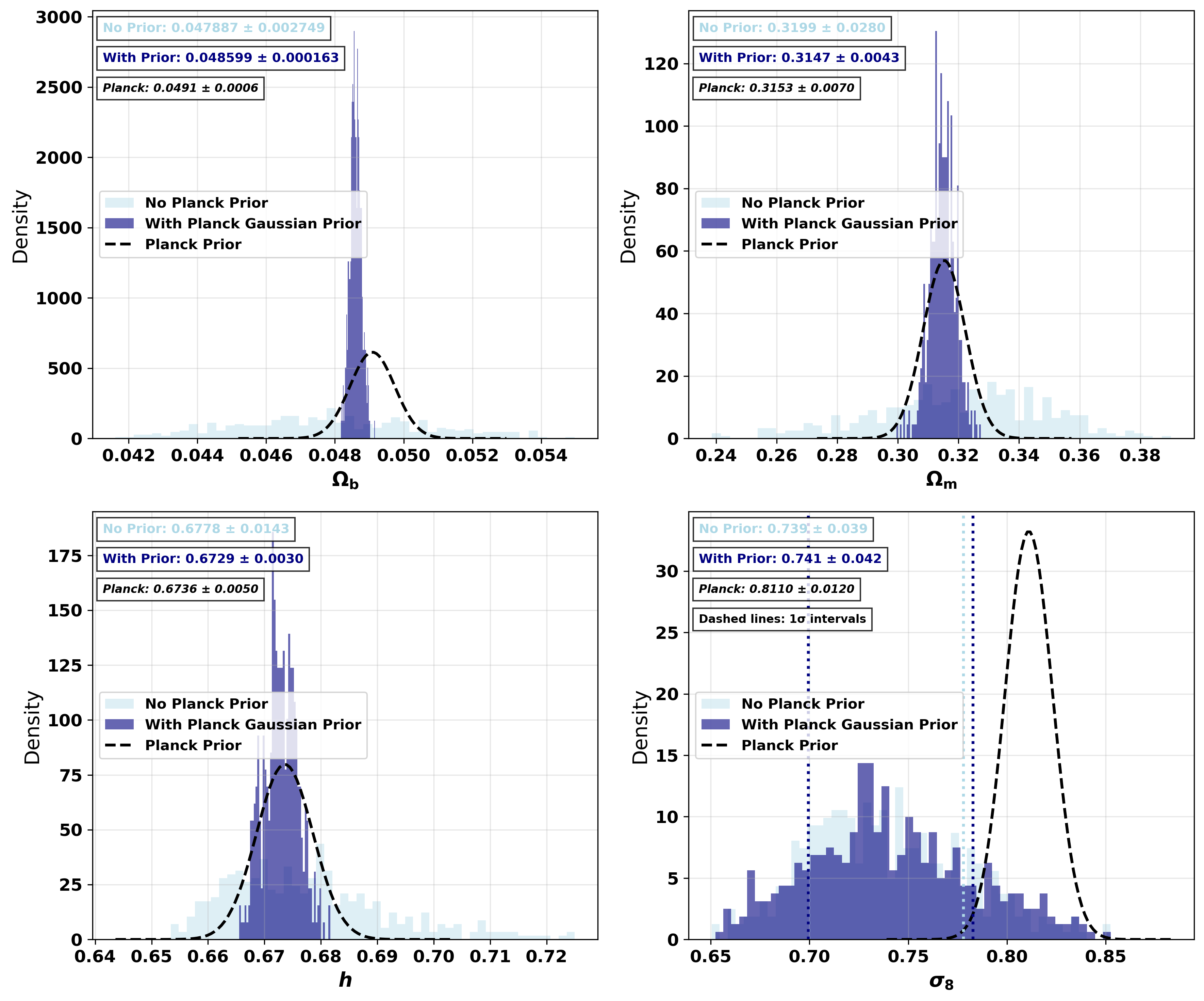}}
    \caption{{
        One-dimensional marginal posterior distributions for $\Lambda$CDM parameters derived from ISW--tSZ data alone, comparing chains run without Planck priors (light blue) and with Planck Gaussian priors (navy). The black dashed line indicates the Planck prior mean for each parameter. For $\Omega_b$, $\Omega_m$, and $h$, the Planck prior significantly tightens the posterior, bringing estimates within $0.35\sigma$ of the Planck mean, with credible intervals reduced by factors of 4 to 9 compared to the no-prior case. In contrast, the posterior for $\sigma_8$ remains largely unaffected by the prior: the mean shifts only slightly ($\Delta \mu = +0.0023$), and the distribution is inconsistent with the Planck value at approximately $1.7\sigma$. Dashed light blue and navy lines in the $\sigma_8$ panel denote the 1$\sigma$ credible intervals for the respective posteriors.}
    }
    \label{fig:planck_prior_comparison}
\end{figure*}

{To test the stability of our inference under external constraints, we repeat the analysis applying Gaussian priors on base cosmological parameters from \textit{Planck} 2018 TT,TE,EE+lowE+lensing results~\cite{Planck2018TT}:}

\begin{align*}
    \Omega_b h^2 &= 0.02237 \pm 0.00015 \\
    \Omega_c h^2 &= 0.1200 \pm 0.0012 \\
    H_0 &= 67.36 \pm 0.54 \, \mathrm{km/s/Mpc} \\
    \sigma_8 &= 0.8110 \pm 0.0120
\end{align*}

{The resulting posterior distributions for $\Omega_b$, $\Omega_m$, $h$, and $\sigma_8$ are compared in Figure~\ref{fig:planck_prior_comparison} between runs with no prior (light blue) and with Planck Gaussian priors (navy). The black dashed line in each panel indicates the mean of the Planck prior.}

\begin{figure*}
    \centerline{\includegraphics[width=15cm]{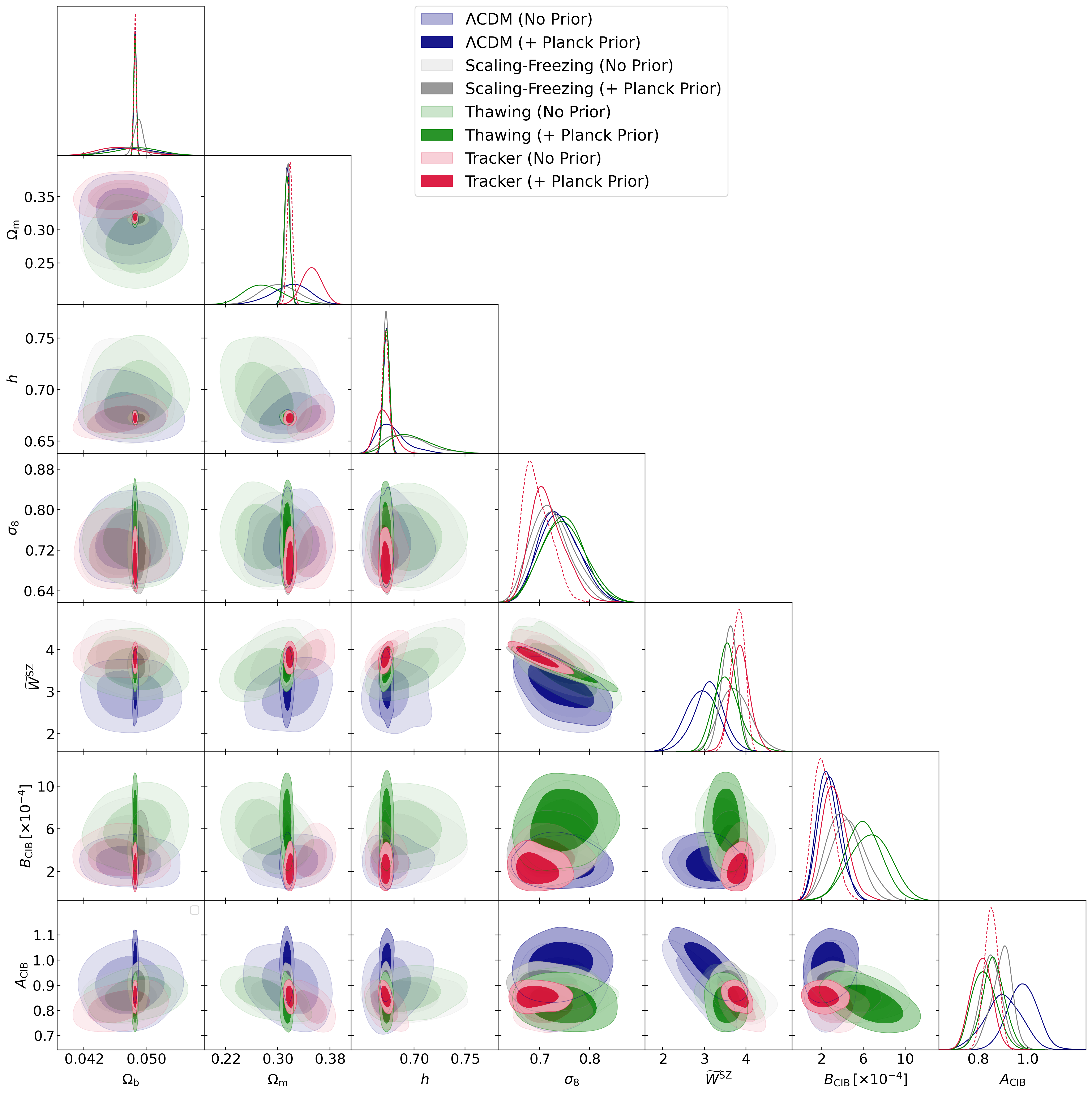}}
    \caption{{
        Triangle plot showing 68\% and 95\% credible regions for cosmological parameters in four models ($\Lambda$CDM, Scaling Freezing, Thawing, Tracker) using ISW--tSZ data alone (lighter contours) and with Planck Gaussian priors (darker contours).
    }}
    \label{fig:all_models_with_without_planck}
\end{figure*}

{As shown in Figure~\ref{fig:planck_prior_comparison}, the inclusion of Planck priors has a clear and differential effect across parameters. For $\Omega_b$ and $h$, the posteriors are pulled toward the Planck values, with the 1$\sigma$ credible intervals narrowing substantially by factors of 4 to 9. This indicates that while ISW--tSZ data alone provides only weak constraints on these parameters, it remains broadly consistent with Planck when combined with external information. The posterior for $\Omega_m$ also tightens and shifts toward the Planck value, though less dramatically than for $\Omega_b$ and $h$, reflecting residual degeneracies with other parameters in the ISW--tSZ likelihood.}

{Critically, the posterior for $\sigma_8$ shows minimal response to the Planck prior: the peak remains near $0.74$, while Planck’s value is $0.811 \pm 0.012$, corresponding to a $\sim 1.7\sigma$ tension that persists even after incorporating the prior. The persistence of the $\sigma_8$ offset unchanged by external priors reflects a difference in probe sensitivity: the ISW--tSZ signal is primarily sensitive to the integrated gravitational potential along the line of sight, not to the present-day amplitude of matter fluctuations.}

{These findings highlight two features of the ISW–tSZ probe. First, it is not a direct probe of $\sigma_8$; its insensitivity to small-scale structure leads to weaker constraints and potential biases in this parameter when used in isolation. Second, the persistent $\sigma_8$ tension underscores the complementary nature of ISW–tSZ: it may probe aspects of late-time cosmology that are not fully captured by standard $\Lambda$CDM fits to CMB data. Resolving this tension will require joint analyses with probes sensitive to small-scale structure, such as galaxy clustering, weak lensing, or cluster counts.}

{To further evaluate the impact of Planck priors on our main conclusion, we performed a full comparison of all four models ($\Lambda$CDM, Thawing, Tracker, Scaling Freezing) under Planck Gaussian priors (Figure~\ref{fig:all_models_with_without_planck}). This plot demonstrates that the Thawing model continues to show the strongest preference over $\Lambda$CDM, with tighter constraints and higher density in regions of parameter space favored by the data. All other models exhibit broader posteriors and less pronounced peaks compared to Thawing, consistent with their higher complexity and lower statistical support.}

{Critically, the $\sigma_8$ tension persists across all models: the posterior remains centered near $0.74$, regardless of whether they are canonical scalar field models or extended versions. This indicates that the tension is not an artifact of the prior, but likely reflects a characteristic of the ISW--tSZ signal. The persistent $\sigma_8$ offset further suggests that the ISW--tSZ signal encodes late-time physics that is not fully aligned with early-universe CMB constraints.}

{The Thawing model remains preferred over $\Lambda$CDM under Planck priors ($\Delta\mathrm{BIC} = -3.83$), and the model-specific parameters show minimal shifts (Appendix~\ref{app:planck_priors_dde}). This indicates that the preference is not merely an artifact of the weak constraints on background cosmological parameters in the baseline analysis.} To assess whether these priors impact the model-specific dynamical parameters of our dark energy frameworks (rather than just shared cosmological parameters), we turn to Appendix~\ref{app:planck_priors_dde}.
%--------------------------
% Appendix 2: Impact of Planck Gaussian Priors on Dynamical Dark Energy Model Parameters
% --------------------------
\section{Internal Consistency Check: Stability of Dynamical Dark Energy Parameters Under External Priors}
\label{app:planck_priors_dde}

{Building on the cosmological parameter stability shown in Appendix~\ref{app:planck_priors_cosmo}, we now examine whether the model-specific parameters of our quintessence frameworks—Thawing ($\lambda$), Tracker ($n$), and Scaling-Freezing ($\lambda_1$) are affected by external constraints.}

\begin{figure*}[htbp]
    \centerline{\includegraphics[width=15cm]{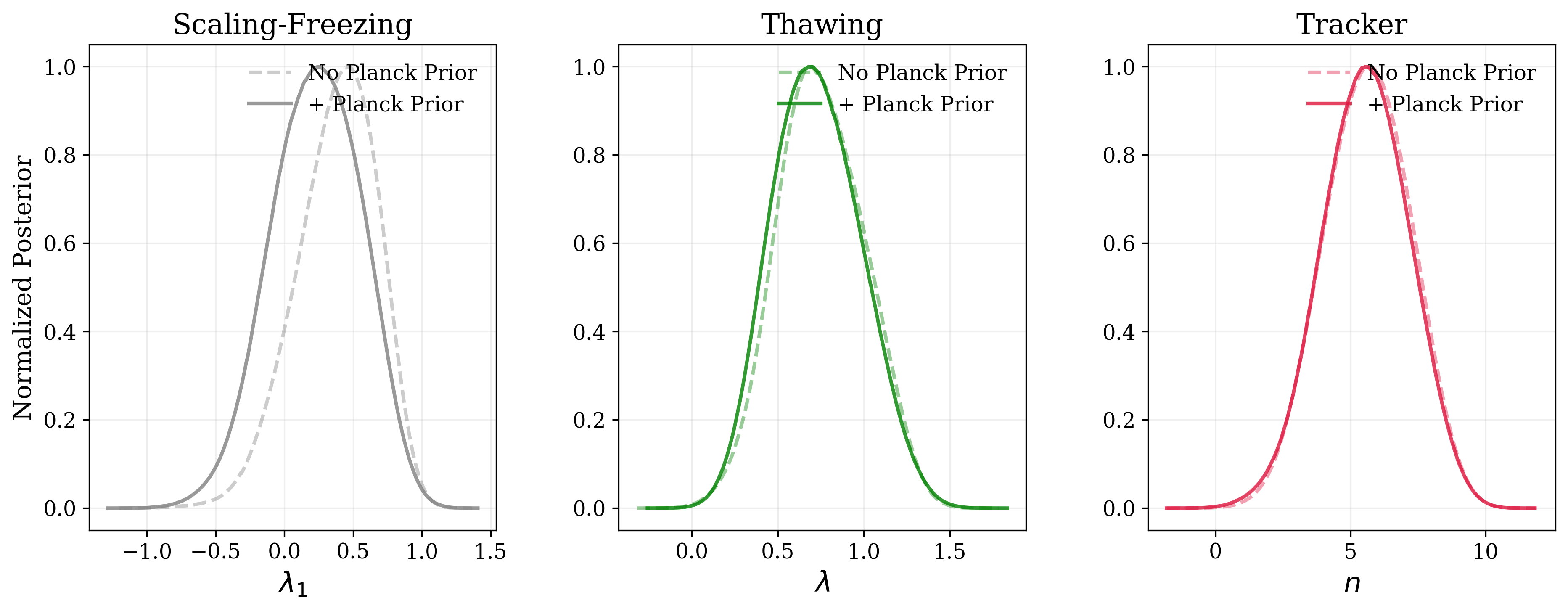}}
    \caption{{
        One-dimensional marginal posterior distributions for model-specific dark energy parameters: Scaling-Freezing $\lambda_1$ (left), Thawing $\lambda$ (middle), and Tracker $n$ (right). Light dashed lines show posteriors from ISW--tSZ data alone (no Planck prior), while dark solid lines show posteriors with Planck Gaussian priors applied. The Planck prior has minimal impact on these parameters: 1$\sigma$ credible interval widths change by only 13.6\% ($\lambda_1$), 6.4\% ($\lambda$), and 0.9\% ($n$)-a stark contrast to the 300–800\% narrowing observed for cosmological parameters like $\Omega_b$. This stability confirms that the inferred dark energy dynamics are weakly degenerate with background cosmology.}}
    \label{fig:model_params_prior_comparison}
\end{figure*}

{As shown in Figure~\ref{fig:model_params_prior_comparison}, the application of Planck Gaussian priors induces negligible shifts in the posterior distributions of model-specific parameters. For Scaling-Freezing $\lambda_1$, the 1$\sigma$ interval changes from $[0.08, 0.70]$ (no prior) to $[-0.10, 0.60]$ (with prior), a width variation of 13.6\%. For Thawing $\lambda$, the interval shifts from $[0.51, 1.01]$ to $[0.46, 0.99]$, a 6.4\% change. For Tracker $n$, the shift is even smaller: from $[4.05, 7.28]$ to $[3.97, 7.22]$, a mere 0.9\% change.}

{This contrasts sharply with cosmological parameters such as $\Omega_b$, whose uncertainties shrink by factors of 4–9 under the same priors. The difference arises because Planck priors constrain early-universe physics, while model-specific parameters govern late-time dark energy behavior, precisely the regime probed by ISW--tSZ.}

{The minimal response of these parameters to external priors shows that the relative ranking of models is not driven by volume effects in the background cosmology. However, it does not imply that the ISW--tSZ data provides definitive evidence for dynamical dark energy. The observed preference for Thawing should be interpreted as internal consistency within our modeling framework, not as confirmation of new physics.}

{Future joint analyses with probes that directly trace the expansion history (e.g., SNe Ia, BAO, weak lensing) will be needed to test whether this preference persists in a broader cosmological context. For this work, which isolates the ISW--tSZ signal as a standalone probe, the stability of model-specific parameters under Planck priors serves as a necessary internal consistency check, ensuring that our conclusions are not artifacts of unconstrained base cosmology.}

%\bibliography{ref_used}
%\bibliographystyle{apsrev4-1}

\end{document}